%% file: main.tex
\documentclass[lettersize,journal]{IEEEtran}

\usepackage{amsmath,amsfonts}
\usepackage{algorithmic}
\usepackage{algorithm}
\usepackage{array}
\usepackage[caption=false,font=normalsize,labelfont=sf,textfont=sf]{subfig}
\usepackage{textcomp}
\usepackage{stfloats}
\usepackage{url}
\usepackage{verbatim}
\usepackage{graphicx}
\usepackage{cite}

\usepackage{pgfplots}
\pgfplotsset{compat=1.18}
\usepgfplotslibrary{groupplots}

\usepackage{tikz}
\usetikzlibrary{arrows.meta,positioning,fit,backgrounds}
\usepackage{xurl}

\input{custom_imports}

\begin{document}


\title{\acronym: Evaluating the Behavioural Correctness of LLM-Generated Smart Contracts}

\author{Giovanni~Pio~Delvecchio$^{\ast}$,
        Stefano~Fantazzini$^{\ast}$,
        Giacomo~Frisoni$^{\ast}$,
        Gianluca~Moro$^{\ast}$,
        Lorenzo~Bacchiani, 
        Mario~Bravetti,
        and~Thomas~Troels~Hildebrandt%
\thanks{Giovanni Pio Delvecchio, Stefano Fantazzini, Giacomo Frisoni, Gianluca Moro, Mario Bravetti, and Lorenzo Bacchiani are with the University of Bologna, Bologna, Italy.}%
\thanks{Thomas Troels Hildebrandt is with the University of Copenhagen, Copenhagen, Denmark.}%
\thanks{$^{\ast}$These authors contributed equally to this work (co-first authorship).}%
\thanks{Manuscript received September 28, 2026.}}


\markboth{Preprint under review}%
{Delvecchio \MakeLowercase{\textit{et al.}}: SolBehave}

\IEEEpubid{\parbox{\textwidth}{\centering This work has been submitted to the IEEE for possible publication.\\ Copyright may be transferred without notice, after which this version may no longer be accessible.}}

\maketitle


\input{sections/abstract}

\input{sections/keywords}
\input{sections/introduction}
\input{sections/preliminaries}
\input{sections/related_work}

\input{sections/methodology}
\input{sections/evaluation}

\input{sections/threats_to_validity}
\input{sections/conclusions}


\bibliographystyle{IEEEtran}
\bibliography{references}

\input{sections/bio}

\end{document}

%% file: custom_imports.tex
\usepackage{comment}

\usepackage{lipsum}

\usepackage{float}

\usepackage{hyperref}

\usepackage{amsfonts}
\usepackage{amssymb}
\usepackage{textcomp}
\usepackage{pifont} 
\usepackage{fontawesome5} 
\usepackage[clock]{ifsym} 
\usepackage[notextcomp]{stix} 
\usepackage{upgreek} 
\usepackage{bbding}
\usepackage{xspace} 
\usepackage{upquote}
\usepackage{verbatim}

\usepackage[table,dvipsnames]{xcolor}
\usepackage{transparent}

\usepackage[most]{tcolorbox} 

\usepackage{adjustbox}

\usepackage{xparse}
\usepackage{scalerel} 
\usepackage{graphicx}
\usepackage{pgfplots}
\usepackage{pgfplots}
\pgfplotsset{compat=1.18} 
\usepackage{pgfplotstable}
\usepgfplotslibrary{groupplots}
\usepgfplotslibrary{statistics}
\usetikzlibrary{matrix} 
\pgfplotsset{compat=newest}
\usetikzlibrary{pgfplots.fillbetween,patterns} 
\usetikzlibrary{shapes.misc}
\usepackage{tikz}
\pgfplotsset{compat=1.18,
    /pgfplots/xbar legend/.style={
    /pgfplots/legend image code/.code={%
       \draw[##1,/tikz/.cd,yshift=-0.25em]
        (0cm,0cm) rectangle (3pt,0.8em);},
   },
   /pgfplots/ybar legend/.style={
    /pgfplots/legend image code/.code={%
       \draw[##1,/tikz/.cd,yshift=-0.25em]
        (0cm,0cm) rectangle (3pt,0.8em);},
   },
}

\usepackage{array}
\usepackage{multirow}
\usepackage{colortbl}
\usepackage{arydshln} 
\usepackage{tabularx} 
\usepackage[para,online,flushleft]{threeparttable} 
\usepackage{stackengine}
\usepackage{hhline} 
\usepackage{diagbox} 
\usepackage{multicol}
\usepackage{booktabs}
\usepackage{makecell} 

\usepackage{bm} 
\usepackage{amsfonts} 
\usepackage{nicefrac} 

\usepackage{mathtools}

\usepackage{enumitem} 

\usepackage[nameinlink,capitalise]{cleveref}
\crefname{section}{\S\hspace{-0.2em}}{\S}

\usepackage{listings}

\definecolor{tablehighlight}{HTML}{F5F5F5}
\definecolor{passgreen}{HTML}{2ECC71}
\definecolor{warnamber}{HTML}{F39C12}
\definecolor{failred}{HTML}{E74C3C}

\definecolor{plotbackground}{HTML}{F4F7FB}

\definecolor{commentgreen}{HTML}{83D475}
\definecolor{functionblue}{HTML}{0000B2}

\definecolor{segO2}{gray}{0.85}
\definecolor{segBoth}{gray}{0.6}
\definecolor{segO3}{gray}{0.25}

\lstdefinelanguage{Solidity}{
  keywords={contract,function,modifier,event,mapping,address,uint256,bool,
             string,bytes,bytes32,returns,return,require,emit,if,else,
             external,internal,virtual,override,payable,view,pure,
             public,private,memory,calldata,storage,unchecked,assembly,
             true,false,revert,for,while,struct,enum,interface,library,
             using,for,import,pragma,solidity,is,new,delete,try,catch},
  sensitive=true,
  morecomment=[l]{//},
  morecomment=[s]{/*}{*/},
  morestring=[b]",
  morestring=[b]',
}

\newcommand{\acronym}{SolBehave}

\newcommand{\yes}{\textcolor{ForestGreen}{\ding{51}}}
\newcommand{\no}{\textcolor{BrickRed}{\ding{55}}}

\newcommand{\dcrcondition}{\mathrel{\rightarrow\!\bullet}}
\newcommand{\dcrresponse}{\mathrel{\bullet\!\rightarrow}}
\newcommand{\dcrinclusion}{\mathrel{\rightarrow\!+}}
\newcommand{\dcrexclusion}{\mathrel{\rightarrow\!\%}}

\newcommand{\iconat}[1]{\raisebox{-0.15ex}{\includegraphics[height=1.6ex,keepaspectratio]{#1}}\,}
\newcommand{\gemmaIcon}{\iconat{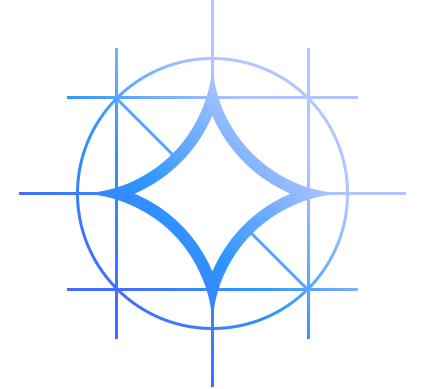}}
\newcommand{\qwenIcon}{\iconat{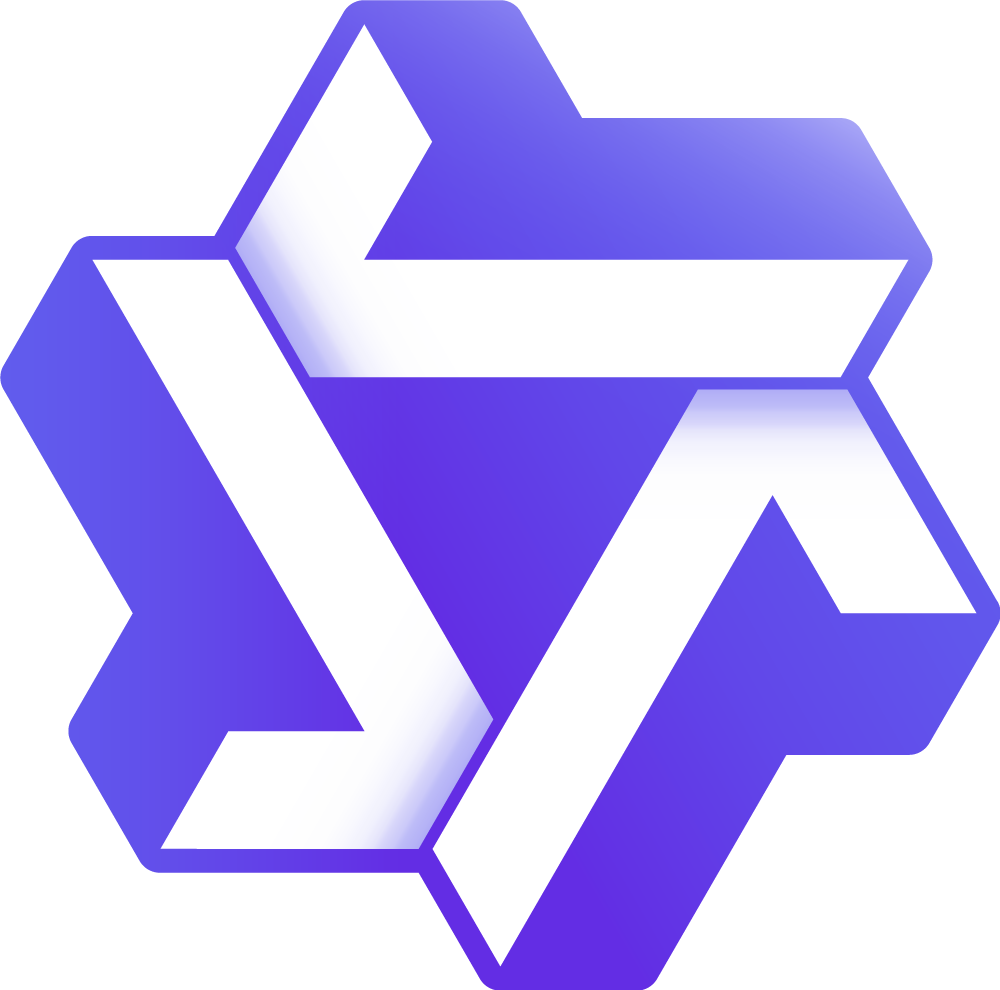}}
\newcommand{\ornithIcon}{\iconat{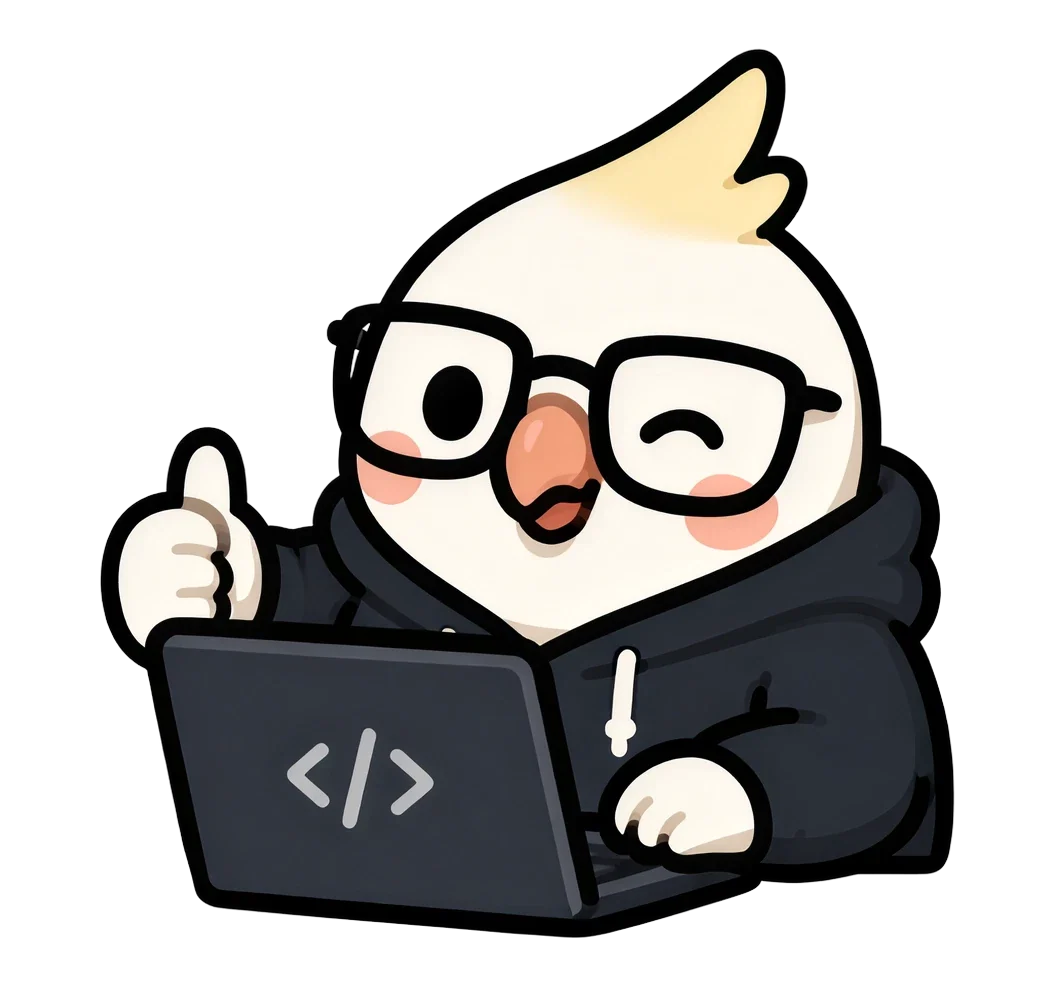}}
\newcommand{\mistralIcon}{\iconat{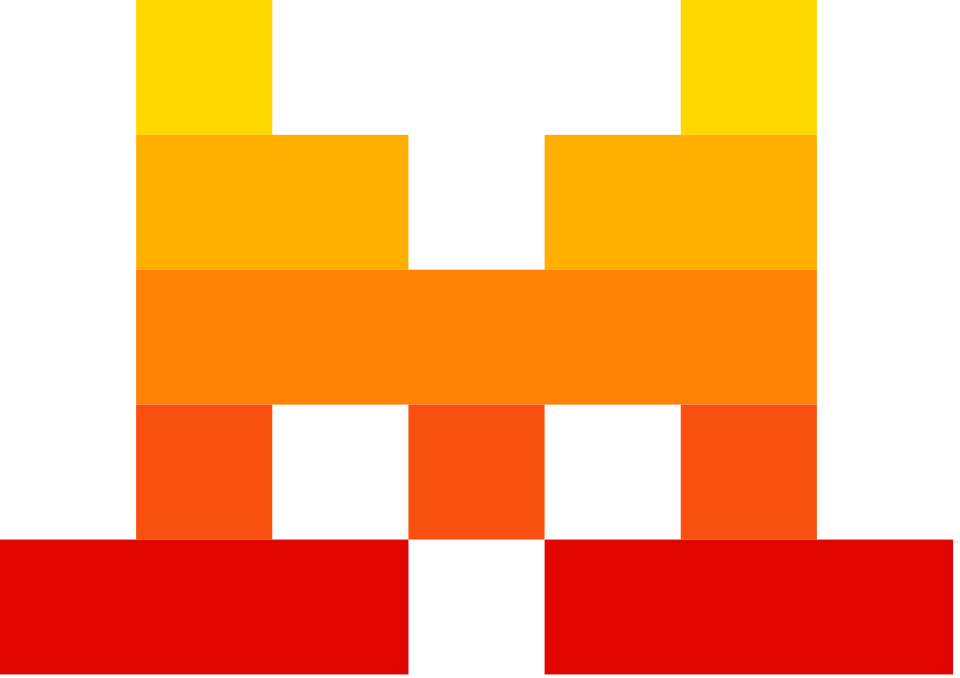}}

\definecolor{gaingrey}{gray}{0.82}
\newcommand{\gaincell}[2]{\cellcolor{gaingrey!#1}$+#2$}   
\newcommand{\hicon}[1]{{#1}\hspace{0.45em}}            
\newcommand{\divhdr}{$\hat{C}\neq C^*$}                
\newcommand{\hstrut}{\rule{0pt}{2.6ex}}                

%% file: sections/abstract.tex
\begin{abstract}
Large language models increasingly generate Solidity smart contracts that, once deployed, can never be patched, making a single latent fault permanent and exploitable.
Yet current evaluations mostly ask only whether generated code compiles.
We introduce \acronym, a benchmark for evaluating the correctness of model-generated contracts at scale, moving beyond ad hoc unit tests.
Across 250 real, reproducible Ethereum contracts, we apply three complementary tests of behavioural fidelity: differential fuzzing, historical-transaction replay, and, as our core contribution, synthetic-transaction replay guided by a Dynamic Condition Response (DCR) graph that enumerates the space of legitimate function-call sequences.
Differential fuzzing compares each generated function with the original on automatically generated inputs; historical replay compares the whole contract on the transactions it actually received on-chain; DCR-guided replay compares it on sequences of calls that follow its intended usage, including orderings that never occurred on-chain.
We evaluate seven state-of-the-art open-weight models after one round of automated repair.
Compilation success ranges from 26.4\% to 92.7\%, but under differential fuzzing only 31.4\% of all generated function bodies that compile, pooled over models and samples, are equivalent to the original.
Divergence at the contract level is the norm rather than the exception: for six of the seven models, between 13\% and 31\% of the generated contracts that compile diverge from the original under historical replay, and between 13\% and 44\% under DCR-guided replay; only Gemma-4-31B stays below 3\% on both.
Compilation is therefore a necessary condition for correctness and far from a sufficient one.
\end{abstract}

%% file: sections/keywords.tex

%% file: sections/introduction.tex
\section{Introduction}
\label{sec:intro}

Smart contracts, programs deployed on a blockchain, increasingly govern real economic activity, and the stakes of getting them wrong have grown with it.
The clearest example is tokenisation, the representation of a real-world or financial asset, such as a bond, a fund share, or a unit of real estate, as a digital token that is transferred and settled directly on-chain, without the intermediaries a traditional financial system relies on.
What began as experimentation has become institutional practice: in 2021 the European Investment Bank issued a 100\,million euro bond on Ethereum~\cite{eib2021bond}, the Bank for International Settlements has since identified tokenisation as a structural shift in how financial markets operate~\cite{bis_tokenisation}, and in 2024 the chief executive of BlackRock, the world's largest asset manager, stated that ``the next generation for markets will be the tokenisation of every financial asset''~\cite{fink2024tokenization,blackrock2024buidl}.
\IEEEpubidadjcol
Nor is finance the only domain: supply chains, for instance, rely on the same guarantees of transparency and auditability, and IBM's Food Trust platform cut Walmart's mango-traceability time from seven days to 2.2\,seconds~\cite{ghosh2026tracking,mohammed2023exploring,joo2021evidence}.
\begin{figure}[!t]
    \centering
    \includegraphics[width=\linewidth]{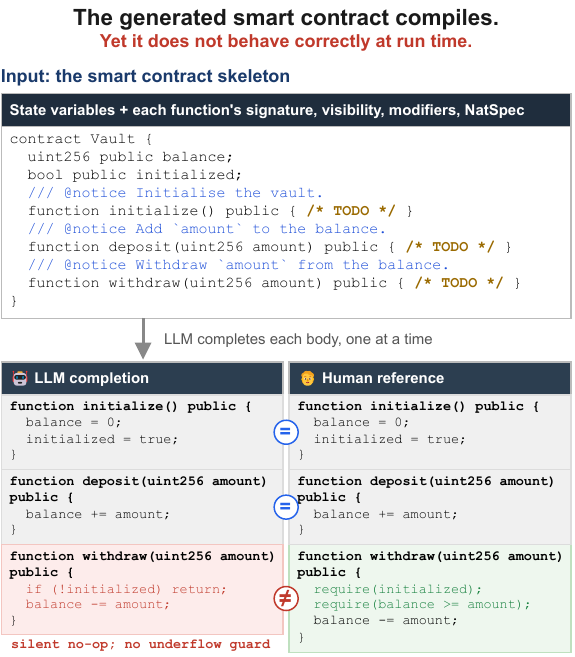}
    \caption{\textbf{Compiling is not behaving.} Given a \texttt{Vault} contract with its function bodies removed (top), a model regenerates them one at a time (bottom left), shown against the original implementation (bottom right). \texttt{initialize} and \texttt{deposit} are regenerated correctly ($=$); \texttt{withdraw} is not ($\neq$). Both versions compile, so compile-rate metrics cannot tell them apart; the behavioural oracles of our benchmark are built to catch exactly this divergence.}
    \label{fig:abstract}
\end{figure}
The contracts behind these systems collectively secure digital assets worth billions of dollars\footnote{\url{https://defillama.com/}}, yet, unlike conventional software, they cannot be patched once deployed.
Any bug present in the bytecode at deployment is permanent and exploitable by anyone, so a single latent fault can cost real money, expose real users, and cannot be undone.

Large Language Models (LLMs) have demonstrated remarkable ability to generate syntactically valid Solidity code~\cite{10.1145/3709016.3737802,DBLP:conf/emnlp/PengYQLLZYL25,DBLP:journals/corr/abs-2503-01098}, and developers are increasingly integrating LLM-based code generation into smart contract workflows~\cite{10.1145/3709016.3737802}.
However, the dominant evaluation paradigm measures compilation success: whether the generated code compiles without errors.
We argue that compilation is a necessary but insufficient filter: necessary because non-compiling code is unusable, but insufficient because syntactically correct Solidity can silently contain dangerous semantic faults that alter token ownership, destroy collected revenue, or enable financial attacks.
Compilation says nothing about whether a contract behaves as intended, and it is precisely this behavioural gap that we target.

Since an arbitrary generated contract comes with no specification of correct behaviour other than the original it was regenerated from, the natural response is differential testing: deploy the generated contract and the original side by side on a sandboxed chain, submit the same calls to both, and flag any divergence in whether a call succeeds or reverts, in what it returns, in the events it emits, or in the storage state it leaves behind.
What distinguishes differential-testing approaches is where the calls come from, and the literature has explored two sources.
The first is historical transaction replay, adopted by SolContractEval~\cite{DBLP:conf/kbse/YeCSBYL25}: the calls are the transactions that real users actually sent to the original contract on mainnet, replayed in their original order, so the test reflects how the contract was really used.
The second is differential fuzzing, adopted by SolBench~\cite{DBLP:journals/corr/abs-2503-01098}: the calls are generated automatically by a fuzzer that searches the input space of each function for arguments on which the two implementations disagree, so the test covers inputs that no user ever sent.
Both are sound and validated notions of behavioural correctness, and we adopt both.
Yet neither can guarantee that the test reaches every state the contract is designed to enter.
Historical replay preserves the order of real calls, but only of the calls that happened to occur: a state that no user ever brought about is never tested.
Differential fuzzing explores each function's inputs in depth, but tests functions in isolation, so a state that can only be reached by combining several calls is never tested either.

Figure~\ref{fig:abstract} shows why this matters, using a toy \texttt{Vault} contract.
The missing balance check in the regenerated \texttt{withdraw} is reachable only after \texttt{initialize} has been called; once the vault is initialised, an over-withdrawal is rejected by the original but executed by the regenerated code, which, in the absence of checked arithmetic, corrupts storage silently.
Historical replay exposes the fault only if some recorded transaction happened to call \texttt{initialize} before an over-withdrawal, and fuzzing tests \texttt{withdraw} on its own.
In both cases the fault goes unreported, not because the two implementations agree, but because the test never reaches the state in which they disagree.

Rather than replacing either approach, this paper unifies them within a single framework and contributes a third, complementary oracle.
We model the contract's intended lifecycle as a \emph{Dynamic Condition Response (DCR) graph}~\cite{DBLP:journals/corr/abs-1110-4161}, a lightweight declarative formalism that captures the condition, response, exclusion, and inclusion relations among contract events.
The graph then \emph{steers} test-sequence generation toward the prerequisite orderings that matter, exercising behaviours that neither historical traffic nor function-level fuzzing reaches.

This work addresses two gaps: the absence of a behavioural benchmark for LLM-generated smart contracts, and the lack of oracles able to detect behavioural divergence from a contract's intended functionality.
Concretely:

\begin{enumerate}[leftmargin=*, itemsep=3pt]

    \item \textbf{A family of three behavioural fidelity oracles} (O1--O3) that complement the standard compile@$k$ pre-filter. O1 applies differential fuzzing, asking whether each regenerated function remains equivalent to the original under coverage-guided input exploration. O2 replays historical mainnet transactions, asking whether the regenerated contract reproduces the behaviour observed on-chain. O3 replays test sequences synthesised from an LLM-extracted DCR graph of the contract's lifecycle, reaching states that historical traffic and unguided fuzzing miss.
    
    \item \textbf{A neuro-symbolic transaction sequence generator} that combines the permissive declarative semantics of DCR graphs with on-chain guard evaluation and LLM-guided parameter synthesis. An LLM extracts the DCR graph from the contract's source, and a deterministic graph-traversal algorithm enumerates legal event orderings and validates each call by executing it on the original contract, so
    correctness follows from the algorithm's guarantees rather than the model's output.
    
    \item \textbf{An open behavioural benchmark and replication package} available at \url{https://github.com/disi-unibo-nlp/sol-behave} (MIT license). It contains the 250 benchmark contracts with their deployment metadata, the DCR graph extracted for each contract, the transaction sequences, and the scripts to regenerate contracts and run the three metrics.

\end{enumerate}

Applying the framework to seven state-of-the-art open-weight models yields three findings.
First, compilation is a poor proxy for correctness: even after automated repair, the models compile between 26.4\% and 92.7\% of their contracts, but only a minority of the regenerated function bodies that compile are equivalent to the original under differential fuzzing (31.4\%, pooled over models and samples), and for six of the seven models at least one regenerated contract in eight diverges from the original under each replay oracle.
Second, the three oracles are complementary rather than redundant: each detects divergent contracts that the others accept as correct, because each exercises the contract differently.
In particular, DCR-guided sequences call functions that real users never did, typically owner-only setup and administrative operations, on one contract in three; this holds regardless of which model extracts the graph.
Third, LLM-extracted DCR graphs are reliable enough to drive the oracle: against human-annotated ground truth over four relation types (\texttt{condition}, \texttt{exclusion}, \texttt{inclusion}, \texttt{response}), extraction reaches 0.72 precision and 0.88 recall overall, and 0.96 precision on \texttt{condition}, the relation that governs call ordering.

\paragraph*{Paper structure}
Section~\ref{sec:preliminaries} introduces the necessary blockchain and smart-contract background.
Section~\ref{sec:related_work} surveys related work.
Section~\ref{sec:methodology} formalises the framework and its oracles.
Section~\ref{sec:evaluation} reports the empirical results.
Section~\ref{sec:threats} discusses threats to validity.
Section~\ref{sec:conclusion} concludes.

%% file: sections/preliminaries.tex
\section{Preliminaries}
\label{sec:preliminaries}

\paragraph{Blockchain}
A shared ledger of transactions, replicated across many independent nodes, that grows only by appending blocks.
Each block contains a cryptographic hash of the previous one, so altering any past transaction would invalidate every block after it.
Once recorded, the history is therefore both permanent and publicly visible, which is what makes it possible to replay a contract's entire past.

\paragraph{Ethereum}
The blockchain platform this work targets~\cite{ethereumdocs}.
It runs programs through the Ethereum Virtual Machine (EVM), where every operation consumes \emph{gas}, the unit in which computation is priced.

\paragraph{Smart contract}
A program stored on the blockchain together with its persistent data, its \emph{storage}.
Its code cannot be changed after deployment, it runs only when a transaction calls one of its public functions, and its execution is deterministic: the same call on the same storage always yields the same result, so any transaction, past (O2) or newly generated (O3), can be re-executed with identical effects.
Only state-changing functions can alter storage; read-only \emph{view} and \emph{pure} functions cannot, and are therefore excluded from behavioural comparison, while the \emph{constructor} runs once at deployment to set the initial storage.

\paragraph{Transaction, revert, and event}
A transaction is a call to a contract function, with given arguments, from a given sender.
It either \emph{succeeds}, making its storage changes permanent, or \emph{reverts}, undoing them, typically because a check such as \texttt{require} failed.
While running, a contract may emit \emph{events}, log entries through which applications outside the chain observe what it did.
Success or revert, return value, emitted events, and resulting storage are the outcomes on which our oracles compare an original contract with a regenerated one.

\paragraph{Solidity}
The most widely used language for EVM contracts and the language of the code we evaluate~\cite{soliditydocs}.
A contract declares state variables and functions, each with a visibility.
A function may carry \emph{modifiers}, reusable checks run before its body (for instance, restricting it to the owner), and NatSpec comments (\texttt{@notice}, \texttt{@param}, \texttt{@return}) documenting it in natural language.
Compiling a contract produces, besides its bytecode, an Application Binary Interface (ABI): the list of its functions with their argument and return types, which tells a caller how to format a call to the contract and how to read the result.

\paragraph{Etherscan}
A public website and API that indexes every Ethereum transaction and hosts the source code of contracts whose deployers submitted it, verified against the deployed bytecode~\cite{etherscandocs}.
We use it for both the past transactions (O2) and the verified source of our benchmark contracts.

\paragraph{Hardhat}
A development environment that runs a private Ethereum node locally~\cite{hardhatdocs}.
We use it to deploy the contracts under test on a clean chain and to execute the transactions of O2 and O3.

\paragraph{Echidna}
A fuzzer for Solidity that generates large numbers of transactions, guided by the code they reach, in search of one that violates a user-specified property, such as two implementations returning different results~\cite{DBLP:conf/issta/GriecoSCFG20}.
It is the engine of our differential-fuzzing oracle (O1).

\paragraph{Tools appearing in related work}
None of the following is part of our pipeline.
\emph{Slither}~\cite{DBLP:conf/icse/FeistGG19} analyses Solidity source without running it, looking for known vulnerability patterns.
\emph{Mythril}~\cite{mueller2018smashing} explores a contract's possible executions symbolically, with an SMT solver~\cite{barrett2018satisfiability}, to find inputs that trigger vulnerabilities.
\emph{Diffusc}~\cite{diffusc} fuzzes two versions of the same contract, before and after an upgrade, in parallel with Echidna and reports behaviours on which they differ.

%% file: sections/related_work.tex
\section{Related Work}
\label{sec:related_work}

\paragraph{Evaluating LLM-generated contracts}
Early evaluations of LLM-generated Solidity rely on static analysers such as Slither and Mythril~\cite{DBLP:conf/compsac/NapoliBGS24,DBLP:conf/icse/DurieuxFAC20,DBLP:conf/issta/RenYMXJSLC21}.
These scale well, but examine the code without running it, so they cannot observe how the generated contract actually behaves.
Execution-based evaluation came later, in the form of differential testing: the generated contract and its original are run on the same inputs and any discrepancy is flagged.
This is also the foundation of our O1 (differential fuzzing) and O2 (historical replay).
SolBench~\cite{DBLP:journals/corr/abs-2503-01098} applies differential fuzzing to 7{,}604 contracts, and SolEval~\cite{DBLP:conf/emnlp/PengYQLLZYL25} complements compile success and pass@k, measured against human-written unit tests, with gas consumption and vulnerability rate.
Both draw their inputs at random or from fixed tests, so neither can guarantee that call sequences requiring specific preconditions are exercised.
SolContractEval~\cite{DBLP:conf/kbse/YeCSBYL25} instead replays real historical transactions against the generated contract.
Replay preserves the original call ordering, but it is bounded by the transaction history itself.
Existing benchmarks thus trade scale for behavioural coverage; Table~\ref{tab:solidity_benchmarks} positions the closest ones along the dimensions that distinguish ours.

\begin{table*}[t]
  \centering
  \caption{\textbf{Comparison of Solidity code-generation benchmarks.}}
  \label{tab:solidity_benchmarks}
  \begin{adjustbox}{width=\linewidth}
  \begin{threeparttable}
  \renewcommand{\arraystretch}{1.45}
  \begin{tabular}{lcccccc}
    \toprule
    \textbf{Benchmark} & \textbf{\#Contracts} & \makecell{\textbf{Contract source}} &
    \multicolumn{1}{c}{\textbf{Evaluation}\tnote{$\spadesuit$}} &
    \makecell{\textbf{Fresh deploy +}\\\textbf{reprod.-verified}\tnote{$\heartsuit$}} &
    \makecell{\textbf{Sequence}\\\textbf{awareness}\tnote{$\clubsuit$}} &
    \makecell{\textbf{Ordering-aware}\\\textbf{coverage}\tnote{$\diamondsuit$}} \\
    \midrule
    Khalid et al.~\cite{10.1145/3709016.3737802} & 102 & GitHub &
      \makecell{Compile@k, Unit-test} & \no & \no & \no \\[3pt]
    SolEval~\cite{DBLP:conf/emnlp/PengYQLLZYL25} & 28 & GitHub &
      \makecell{Pass@k, Compile@k,\\ Gas, Vuln.\ rate} & \no & \no & \no \\[3pt]
    SolContractEval~\cite{DBLP:conf/kbse/YeCSBYL25} & 124 & Etherscan &
      \makecell{Hist.\ trans.\ replay} & \no & \yes & \no \\[3pt]
    SolBench~\cite{DBLP:journals/corr/abs-2503-01098} & 7{,}604 & Etherscan &
      \makecell{Differential fuzzing} & \no & \no & \no \\
    \midrule
    \rowcolor{tablehighlight}
    \textbf{SolBehave (Ours)} & \textbf{250} & Etherscan &
      \makecell{Compile@k, Differential fuzzing,\\ Hist.\ trans.\ replay, DCR replay} & \yes & \yes & \yes \\
    \bottomrule
  \end{tabular}
  \begin{tablenotes}[flushleft]\footnotesize
    \item[$\spadesuit$] \textbf{Gas}: average gas per call. \textbf{Vuln.\ rate}: fraction of generated contracts flagged by static analysers (e.g.,\ Slither).\\
    \item[$\heartsuit$] \textbf{Fresh deploy + reprod.-verified}: the original contract is redeployed from its source code onto a clean blockchain and confirmed to reproduce its own past on-chain behaviour, rather than being tested against a live fork of the chain.\\
    \item[$\clubsuit$] \textbf{Sequence awareness}: functions are called in an order that respects preconditions (e.g.,\ initialising $\rightarrow$ withdrawing), not in isolation.\\
    \item[$\diamondsuit$] \textbf{Ordering-aware coverage}: the evaluation deliberately reaches code paths that are unlocked only by a specific prior call sequence, which real transaction histories rarely exercise.
  \end{tablenotes}
  \end{threeparttable}
  \end{adjustbox}
\end{table*}

\paragraph{Testing hand-written contracts}
A substantial body of work tests \emph{hand-written} contracts through fuzzing~\cite{DBLP:conf/issta/GriecoSCFG20,DBLP:conf/issta/ShouTS23}, symbolic execution~\cite{DBLP:conf/kbse/MossbergMHGGFBD19,mueller2018smashing}, static analysis~\cite{DBLP:conf/icse/FeistGG19}, and formal verification~\cite{jackson2022certora}; LLM4FUZZ~\cite{DBLP:journals/corr/abs-2401-11108} additionally steers a fuzzer with correctness conditions inferred by an LLM.
All of these tools check a contract against properties stated explicitly by the developer, such as an invariant, a vulnerability pattern, or a formal specification, and take the code itself as the intended program.
For generated code, there is no such list of properties: the contract can deviate from the intended behavior in any way.

\paragraph{Sequence-aware fuzzing}
Differential fuzzing, as used in SolBench and in our O1, varies the arguments of one function at a time.
A separate line of bug-finding fuzzers for hand-written contracts varies both: each test is a sequence of several calls, and the fuzzer searches over the order of the calls as well as over their arguments, because some branches can only be reached after other functions have modified the state.
IR-Fuzz~\cite{liu2023rethinking} infers the order from the code: if a function reads a state variable that another function writes, the writer is scheduled first.
MuFuzz~\cite{qian2024mufuzz} instead tracks which sequences bring execution close to an unexplored branch and mutates those, rather than random ones, to reach it.
Both report substantial branch-coverage gains over order-agnostic fuzzers, confirming that call ordering is a first-order factor in exploring a contract's state.
We share the observation that call order matters, but we use it for a different purpose.
These fuzzers look for bugs in a single contract, so they want sequences that reach as much of its code as possible, and they obtain them by analysing the code itself.
We compare two contracts, so we want sequences that the original contract accepts as legitimate uses; only then can a difference in behaviour be blamed on the generated contract.
Such sequences come from a description of how the contract is meant to be used, its DCR graph, rather than from an analysis of what its code happens to do.

\paragraph{Neuro-symbolic program analysis}
Our sequence generator is neuro-symbolic: a neural component, the LLM, produces an artefact that a symbolic component, the graph-traversal algorithm, then executes with formal guarantees.
In the taxonomy of~\cite{delvecchio2025neuro}, what distinguishes such systems is what the LLM produces.
Ours produces a set of constraints that are already implicit in the contract, formalised as the relations of a DCR graph, which is an instance of \emph{rule mining}; by contrast, in program-synthesis systems such as LINC~\cite{olausson2023linc}, the LLM writes an executable logic program that a solver then runs.

\paragraph{DCR graphs}
Introduced by Hildebrandt and Mukkamala~\cite{DBLP:journals/corr/abs-1110-4161} as a declarative, event-based model of loosely constrained processes, DCR graphs have been used to capture smart contract design patterns~\cite{DBLP:conf/sefm/EshghieAAHS23} and, in HighGuard~\cite{DBLP:conf/kbse/EshghieASAHS24}, to monitor contract executions at run time against a developer-written specification.
We differ on both counts: our graph is extracted automatically from the source rather than written by hand, and it is used proactively to generate transactions rather than to judge them.
To our knowledge, this is the first use of DCR graphs for test-sequence generation in the evaluation of LLM-generated code.

%% file: sections/methodology.tex
\section{Methodology}
\label{sec:methodology}

Our framework proceeds in two stages.
First, an LLM regenerates each benchmark contract (\S\ref{sec:codegen}).
Second, the regenerated samples are evaluated against the deployed original through a syntactic filter, Compile@$k$, and three behavioural oracles, namely differential fuzzing (O1, \S\ref{sec:o1}), historical transaction replay (O2, \S\ref{sec:o2}), and DCR-guided synthetic replay (O3, \S\ref{sec:o3}).
Throughout, $C^*$ denotes the original contract, taken as ground truth, and $\hat{C}$ a regenerated sample.
Both replay oracles compare $C^*$ and $\hat{C}$ on the same quantities, defined in \S\ref{sec:comparison-axes}.
We first describe the set of contracts to which we apply the framework (\S\ref{sec:benchmark}).

\subsection{Benchmark Construction}
\label{sec:benchmark}

We build the benchmark in three steps: crawling candidate contracts, keeping those that reproduce their own on-chain history on a fresh deployment, and restricting the set to contracts of a size the generation task can handle.

\paragraph{Crawling}
We crawl Ethereum mainnet from block 16{,}308{,}190 (2023-01-01) onward, sampling blocks in batches and ranking call targets by transaction frequency.
A candidate is kept only if it exposes verified source on Etherscan, is not a proxy, and has accumulated at least 200 historical transactions. This yields 8{,}360 candidates.

\paragraph{Reproducibility check}
Each candidate is compiled from its verified source and deployed fresh on a local Hardhat node; constructor arguments are taken from Etherscan's record, decoded from the deployment transaction, or defaulted.
We then replay the contract's first 200 historical transactions against this fresh deployment and compute the fraction that behave as recorded.
Contracts that reproduce less than 80\% of their own recorded history are discarded, leaving 2{,}098.

\paragraph{Size filter}
We count the non-trivial functions of each contract, excluding constructors and \texttt{view}/\texttt{pure} functions, and retain only contracts with between 1 and 20 of them.
Figure~\ref{fig:function-buckets} shows the effect of this filter at the crawled and the reproducibility-verified stage: contracts with more than 20 functions dominate both sets, and the 1--20 band isolates exactly the 250 contracts of our final benchmark.
The bound keeps the generation task tractable while preserving a meaningful range of complexity (median 16 functions, $p_{75}=18$); \S\ref{sec:threats} reports the ablation on which it is based.

\input{figures/function_bucket_distribution}

\subsection{Code Generation}
\label{sec:codegen}
%
Each of the 250 contracts is regenerated from its verified mainnet source by a stub-and-complete pipeline, following the protocol of SolBench~\cite{DBLP:journals/corr/abs-2503-01098}, illustrated in Figure~\ref{fig:abstract}.
All original comments are stripped and replaced by freshly generated NatSpec for every public function, so that no sample benefits from leaked documentation.
The function bodies are then deleted, leaving the signature, visibility, modifiers, and NatSpec, and the model completes the stub one function at a time in source order.
Later functions therefore condition on the code already generated for earlier ones: a function calling an internal helper sees that helper's actual implementation rather than its stub.
We repeat this procedure ten times per contract, independently and from the same skeleton, obtaining $k=10$ samples that share every signature and differ only in their function bodies.
The samples are compiled separately, and we report the fraction that compile, $\mathrm{Compile}@k = c/k$ with $c$ the number of compiling samples.

\subsection{O1: Differential Fuzzing}
\label{sec:o1}

O1 tests whether each regenerated function behaves as the original on arbitrary inputs.
We consider a function \emph{fuzzable} when it changes state and all its arguments are of primitive type.
For every compiling sample and every fuzzable function $f$, we generate a harness contract that deploys $C^*$ and $\hat{C}$ with identical constructor arguments and exposes a wrapper for $f$; the wrapper invokes $f$ on both deployments with the same arguments and asserts that the two calls agree on status and, when both succeed, on the returned data.
Since the harness is itself a contract, it receives the return values, and since it deploys both contracts, it is their owner, so owner-only functions are exercised as well.
Echidna~\cite{DBLP:conf/issta/GriecoSCFG20} searches for arguments that violate the assertion (10{,}000 calls, sequence length 20, 120\,s timeout, following~\cite{DBLP:journals/corr/abs-2503-01098}); no LLM is used.
A regenerated function is \emph{equivalent} to the original if none is found, and O1 reports the fraction of equivalent functions.
Each wrapper targets a single function, so O1 accumulates state across repeated calls to that function, but never interleaves different functions.
Our numbers are not comparable with those of SolBench~\cite{DBLP:journals/corr/abs-2503-01098}, which fuzzes each generated function inside an otherwise original contract, so that no other function can be at fault.
We fuzz inside a fully regenerated contract, as a developer would deploy it, where a fault in one function can corrupt the state that another one reads.

\subsection{O2: Historical Transaction Replay}
\label{sec:o2}

O2 replays each contract's own recorded mainnet history against a freshly deployed instance.
We take the first $1{,}000$ historical transactions and deploy both $C^*$ and $\hat{C}$ from source, never by injecting bytecode, which would leave the original deployer as owner and desynchronise every privileged call.
The transactions are then re-issued against both contracts in their original order and compared along the axes of \S\ref{sec:comparison-axes}.
Nothing here is synthesised: the arguments and call order are taken from transactions submitted by real users, which is the strength of this oracle and, as \S\ref{sec:o3} discusses, also its limit.

\subsection{O3: DCR-Guided Synthetic Transaction Replay}
\label{sec:o3}

Instead of depending on the orderings that real users happened to exercise, O3 extracts an explicit model of the contract's intended call structure and synthesises sequences that reach the orderings and preconditions the historical record may never have covered.

\subsubsection{The DCR formalism}
\label{sec:dcr-formalism}

A \emph{DCR graph}~\cite{DBLP:journals/corr/abs-1110-4161} is a declarative process model that specifies which events may occur at each point and how the occurrence of one affects the others.
We adopt the core formalism, since the extensions proposed since its introduction, such as time, data, and nesting, are not required for our purposes.
DCR is deliberately \emph{permissive}: it constrains only the relationships that must hold and leaves every other ordering free.
This suits extraction by an LLM, which need only identify the few constraints that matter without over-constraining the process, and functions with no ordering requirement simply appear as isolated nodes, callable at any time.
A graph consists of \emph{events}, one per public state-changing function branch, each labelled with the function name and a caller role (\texttt{owner} or \texttt{any}).
Every event is at all times either \emph{included} or \emph{excluded}: an included event can be executed and the constraints it places on other events apply, whereas an excluded event is ignored as if it were absent from the graph, so it cannot be executed, it no longer blocks the events it conditions, and any obligation on it lapses.
The state of the graph, its \emph{marking}, is therefore given by which events have been executed, which are currently included, and which are \emph{pending}, that is, obliged to be executed but not yet.
Events are related by four typed \emph{relations}:
\begin{description}[leftmargin=1.2em, itemsep=1pt, parsep=0pt, topsep=2pt]
  \item[Condition, $A \dcrcondition B$:] if $A$ is currently included, $B$ is not enabled until $A$ has been executed.
  \item[Response, $A \dcrresponse B$:] executing $A$ makes $B$ \emph{pending}: $B$ must be executed at some later point for the process to be complete, unless it is excluded first.
  \item[Inclusion, $A \dcrinclusion B$:] executing $A$ includes $B$, so that $B$ becomes executable again and its constraints and obligations apply once more.
  \item[Exclusion, $A \dcrexclusion B$:] executing $A$ excludes $B$, so that $B$ can no longer be executed and the constraints and obligations attached to it are dropped.
\end{description}

\noindent
We extend this core formalism in one respect.
In DCR, a relation always applies; in a smart contract, whether it does can depend on the contract's state.
We therefore allow a relation to carry a \emph{guard}, a Solidity expression on the contract's state variables, and treat the relation as present only while the guard holds.
One modelling convention is also worth noting.
DCR has no primitive for an event that may be executed only once: this is modelled by an exclusion from the event to itself, which excludes it the moment it is executed.
Since such self-exclusions are frequent and mechanical, our extraction schema records them as a flag on the event rather than asking the model to emit them as relations.

\paragraph{Extraction}
The graph is extracted from the original contract, which is the specification the regenerated ones are judged against.
We prompt an LLM with the contract's source and ABI and ask for the graph as JSON.
The model emits one event per public state-changing function branch, and infers the relations between them, with their guards, from the preconditions that gate each function.
For every event, it also proposes an argument value, later used to issue the call, and records the value of any compile-time constant a guard refers to, so that the guard can be evaluated during sequence generation.
Because DCR under-constrains by design, the extractor emits few relations relative to the number of functions, $3.7$ per graph on average across our benchmark.
The prompts, the JSON schema, and the extracted graphs are available in the replication package.
Figure~\ref{fig:dcr-example} shows part of the graph extracted for \textsc{Gifto}\footnote{\href{https://etherscan.io/token/0xC5bBaE50781Be1669306b9e001EFF57a2957b09d}{etherscan.io/token/0xC5bBaE...957b09d}}, selected to exhibit all four relation types.
The contract keeps a record of how many units of a digital currency each address owns, and it also handled the initial sale of that currency: while the sale was open, anyone could send money to the contract and receive units in return, at a fixed price.
The sale is controlled by the address that deployed the contract, its \emph{owner}: \texttt{enableSale} and \texttt{disableSale} include and exclude \texttt{saleContinues}, the event of a successful purchase, so buying is possible only between the two.
The money paid by buyers accumulates in the contract's own balance; \texttt{withdrawEth} transfers it to the owner's address, and only the owner can call it.
A purchase therefore makes \texttt{withdrawEth} pending, under the guard \texttt{balance > 0}: there are proceeds to collect, unless the contract holds none.
The condition from \texttt{setAllowance} to \texttt{transferFromSuccess} covers a separate mechanism of the token: holders can let another address spend part of their tokens, and the transfer is enabled only once that permission has been given.
Finally, \texttt{transferFail}, the branch taken when a transfer is refused, is isolated: it can occur at any time and leaves everything else unchanged.
%
\input{figures/dcr_graph_example}

\subsubsection{Validated Sequence Generation}
\label{sec:seq-gen}

Given a graph and a deployed instance of the original contract, we generate \emph{validated sequences}: ordered lists of calls, each with concrete argument values, that respect the constraints of the graph and have all been executed successfully on a local Hardhat node.
Algorithm~\ref{alg:seqgen} summarises the procedure, which we describe in the order in which it runs.

\begin{algorithm}[t]
\caption{Validated sequence generation for one contract}
\label{alg:seqgen}
\begin{algorithmic}[1]
\REQUIRE DCR graph $G$ of the original contract $C^*$; budget $T$ of successful calls
\ENSURE set of validated sequences $\mathcal{D}$
\STATE $\mathcal{D} \leftarrow \emptyset$; $n \leftarrow 0$
\WHILE{$n < T$}
    \STATE deploy $C^*$ fresh; reset the marking of $G$; $\sigma \leftarrow \langle\rangle$
    \LOOP
        \STATE evaluate the guards of $G$ on the current state of $C^*$
        \STATE $E \leftarrow$ events allowed by the marking of $G$ under the relations whose guards hold
        \STATE remove ownership-transferring events from $E$
        \IF{$E = \emptyset$ \textbf{or} $n = T$} \STATE \textbf{break} \ENDIF
        \STATE $e \leftarrow \textsc{Select}(E)$ $\triangleright$ obligation, prerequisite, or least executed (exploration)
        \STATE execute $e$ on $C^*$ with its extracted arguments
        \IF{the call succeeded}
            \STATE append $e$ to $\sigma$; update the marking of $G$; $n \leftarrow n + 1$
        \ENDIF
    \ENDLOOP
    \STATE execute the ownership-transferring events; append successes to $\sigma$
    \STATE $\mathcal{D} \leftarrow \mathcal{D} \cup \{\sigma\}$
\ENDWHILE
\RETURN $\mathcal{D}$
\end{algorithmic}
\end{algorithm}

\paragraph{Sequence construction}
The procedure collects calls until a budget $T$ of successful ones is reached, building one sequence at a time.
Each sequence starts from a fresh deployment of the contract and the initial marking of the graph, and is extended one call at a time, each time determining which events the graph currently allows and picking one of them; when no event can be executed, the sequence ends and a new one is started from a fresh deployment.

\paragraph{On-chain guard evaluation}
Which relations currently apply, and hence which events are allowed, depends on the guards, and a guard such as \texttt{balance > 0} depends on the contract's state.
Before each step, the procedure reads the value of every variable a guard refers to directly from the deployed contract, through the corresponding getter, and evaluates the guards on those values.
This avoids having to predict how each call changes the state, and guarantees that the values used are the ones the contract really holds.
Only events whose guards hold remain candidates for the selection described next.

\paragraph{Event selection}
The event to execute is chosen among the allowed ones according to a fixed priority.
Highest priority goes to \emph{obligations}: events that a response from an already executed event requires to happen.
Next come \emph{prerequisites}: events that other events are waiting for.
The remaining allowed events are chosen last, by an upper-confidence-bound rule over the attempts and successes recorded for each event during the run on that contract: events not yet attempted come first, then those with the highest success rate, with a bonus for events attempted rarely, so that no part of the contract is left unexercised.
The first two priorities are applied with probability $\beta = 0.7$ and $\alpha = 0.8$ respectively, rather than deterministically, to preserve a degree of exploration beyond the constraints encoded in the graph.
Both values were selected after preliminary tests as a balance between constraint-driven progress and exploration.

\paragraph{Ownership-transferring events are executed last}
Some functions, such as \texttt{renounceOwnership} or \texttt{transferOwnership}, permanently take the owner role away from the address that deployed the contract, which is the address our procedure calls from.
If such a function were executed in the middle of a sequence, every later call reserved to the owner would be rejected, and the rest of the sequence would be lost.
These events are therefore excluded from selection during the $T$ steps and executed at the very end, once the rest of the sequence has been built.
Few contracts are affected: on the graphs extracted by \textsc{Gemma-4-31B}, the strongest extractor in our experiments (\S\ref{sec:evaluation}), ownership-transferring events amount to $0.52$ per contract on average, against $6.0$ events per contract overall, and $71.6\%$ of the contracts have none.

\paragraph{Only admissible calls are retained}
The chosen event is \emph{executed on the original contract} with the argument values that the LLM proposed for it when extracting the graph; if the call reverts, it is discarded and the marking is left unchanged, so a sequence contains only calls the original accepts.
The DCR graph describes how the contract is meant to be used, and the sequences are meant to exercise exactly that: legitimate calls, in a legitimate order, whose outcome on the original is by definition the correct one.
A call the original rejects does not belong to this intended use and is therefore left out; for a call the original accepts, any different outcome on the regenerated contract can only be the regenerated contract's fault.
Across the 250 contracts, Algorithm~\ref{alg:seqgen} attempted $256{,}923$ calls and kept $212{,}904$; the remaining $44{,}019$ ($17.1\%$) reverted and were discarded.
The discard rate varies widely from contract to contract: its median is $4.9\%$, but for 37 contracts it exceeds $98\%$, and for 32 of them it reaches $100\%$, leaving those contracts with no sequence at all.
The cause is the same in every case: since we do not search for new arguments when a guard fails ($0.9\%$ of the kept calls fall back on arguments decoded from historical transactions, the rest use the LLM's), a contract whose guards the extracted arguments can never satisfy loses most or all of its calls.
Pairing the orderings of the DCR graph with a fuzzer for the arguments is the clearest direction for future work.

\subsection{Comparison Axes}
\label{sec:comparison-axes}

All three oracles decide divergence on the same criterion, the \emph{status} of a call: whether it succeeds on both deployments or reverts on both.
O1 also compares the value returned by successful calls, which its harness receives as the calling contract (\S\ref{sec:o1}).
O2 and O3 cannot: they replay transactions sent by user accounts, and Ethereum does not deliver return values to user accounts.
What a user receives is the transaction's receipt, which records status, gas used, and emitted events, and it is this receipt that the replay oracles compare, with status as the criterion and gas and events as secondary observations.
The rest of this section concerns O2 and O3, which differ only in where their transactions come from: each transaction is issued, in the same order and from the same sender, to $C^*$ and to every sample $\hat{C}$, and its outcome on the sample is compared with its outcome on $C^*$ call by call.

\paragraph{Divergence criterion}
The fraction of calls on which a sample agrees with $C^*$ on status is its \emph{status match rate}, and a contract is \emph{divergent} under an oracle if at least one of its samples disagrees with $C^*$ on at least one call.

\paragraph{Secondary observations}
For every call we also record the other fields of the receipt, released with the replication package and used to check the status results rather than as criteria.
\emph{Gas}: the gas used by $\hat{C}$ relative to $C^*$; two implementations of the same logic legitimately differ by a few percent, so only deviations beyond $\pm10\%$ are flagged, as they usually indicate different logic.
\emph{Events}: the number of events the call emits and, for each, its topics, that is, the event signature and indexed arguments, so that the same occurrences are reported with the same key values.
In addition, we periodically read the values returned by the contract's zero-argument \texttt{view} functions, as a snapshot of the state it exposes.
On our benchmark, none of these observations exposed a divergence that status had not (\S\ref{sec:threats}).

\paragraph{Aggregation}
The status match rate of a contract is the mean over its $k$ samples; we also keep the \emph{minimum}, since a fault exposed by one sample can be masked by the mean over ten.
O2 can be applied to every contract with at least one compiling sample, whereas O3 also requires at least one validated sequence, which exists for 218 of the 250 contracts (\S\ref{sec:seq-gen}); \S\ref{sec:evaluation} therefore reports, for each oracle and model, the number of contracts actually compared, which is the denominator of its divergence figures.

%% file: figures/function_bucket_distribution.tex
%
\definecolor{stagecrawled}{HTML}{C8C8C8}
\definecolor{stageverified}{HTML}{4A4A4A}
\definecolor{keptband}{HTML}{EFEFEF}

\begin{figure}[t]
\centering
\begin{tikzpicture}
\begin{axis}[
    ybar,
    bar width=9pt,
    width=\linewidth,
    height=4.6cm,
    ymin=0, ymax=60,
    enlarge x limits=0.13,
    axis lines*=left,
    axis on top,
    symbolic x coords={0,1--5,6--10,11--20,21--30,31+},
    xtick=data,
    xlabel={Non-trivial functions per contract},
    ylabel={Share of contracts (\%)},
    x tick label style={font=\small},
    y tick label style={font=\small},
    label style={font=\small},
    tick align=outside,
    ytick={0,20,40,60},
    nodes near coords,
    every node near coord/.append style={font=\scriptsize, /pgf/number format/.cd, fixed, precision=0},
    clip=false,
]
\draw[fill=keptband, draw=none]
    ([xshift=-16pt]axis cs:1--5,0) rectangle ([xshift=16pt]axis cs:11--20,60);
\node[font=\small, anchor=south] at ([yshift=1pt]axis cs:6--10,60)
    {kept: 250 contracts};

\addplot[fill=stagecrawled, draw=none] coordinates {
    (0,8.99) (1--5,8.83) (6--10,7.80) (11--20,14.22) (21--30,22.03) (31+,38.15)
};
\addplot[fill=stageverified, draw=none] coordinates {
    (0,0.86) (1--5,1.76) (6--10,1.91) (11--20,8.25) (21--30,50.24) (31+,36.99)
};
\end{axis}
\end{tikzpicture}

\vspace{2pt}
{\footnotesize
\textcolor{stagecrawled}{\rule{7pt}{7pt}}~Crawled ($n=8{,}360$)
\hspace{1.5em}
\textcolor{stageverified}{\rule{7pt}{7pt}}~Reproducibility-verified ($n=2{,}098$)
}

\caption{\textbf{Distribution of non-trivial functions per contract at the two stages of benchmark construction.} Large contracts dominate both the crawled set and the subset that passes the reproducibility check; the size filter retains the shaded band, 1 to 20 functions, which contains the 250 contracts of the final benchmark (37, 40, and 173 in its three buckets).}
\label{fig:function-buckets}
\end{figure}
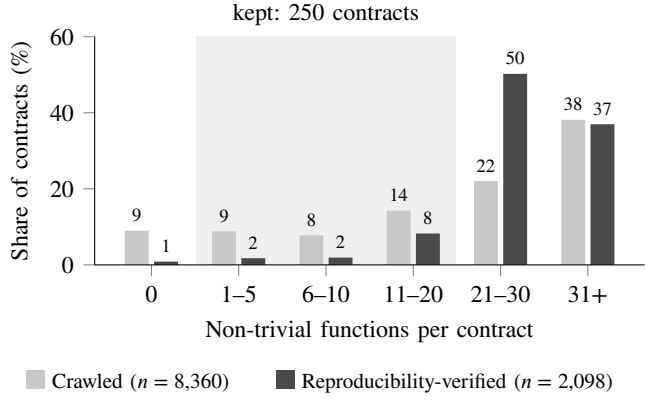

%% file: figures/dcr_graph_example.tex
%
\definecolor{lanefill}{HTML}{FCFCFB}
\definecolor{dcrcondition}{HTML}{345DB7}
\definecolor{dcrresponse}{HTML}{1B8A6B}
\definecolor{dcrinclusion}{HTML}{8156B8}
\definecolor{dcrexclusion}{HTML}{B63F48}

\tikzset{
  event/.style={draw, rounded corners=2pt, minimum width=2.3cm, minimum height=0.8cm,
                align=center, fill=white, thick, font=\footnotesize},
  isolated/.style={event, dotted},
  guard/.style={font=\footnotesize\itshape, fill=white, inner sep=1.5pt},
  lane/.style={draw=black!50, fill=lanefill, dashed, rounded corners=4pt, inner sep=10pt},
  lanelabel/.style={font=\footnotesize\bfseries, anchor=south west, yshift=1pt},
  relcond/.style={line width=1pt, dcrcondition, -{Circle[length=5pt, width=5pt]}},
  relresp/.style={line width=1pt, dcrresponse, {Circle[length=5pt, width=5pt]}-{Stealth[length=2.2mm]}},
  relincl/.style={line width=1pt, dcrinclusion, -{Stealth[length=2.2mm]}},
  relexcl/.style={line width=1pt, dcrexclusion, -{Stealth[length=2.2mm]}},
}

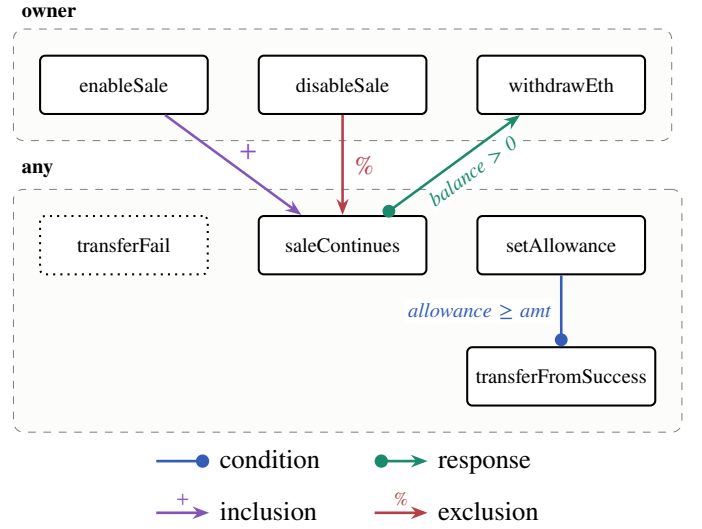
\begin{figure}[t]
\centering
\resizebox{\linewidth}{!}{%
\begin{tikzpicture}
\node[event] (enableSale)  at (1.6,5.0) {enableSale};
\node[event] (disableSale) at (4.6,5.0) {disableSale};
\node[event] (withdrawEth) at (7.6,5.0) {withdrawEth};
\node[isolated] (transferFail) at (1.6,2.8) {transferFail};
\node[event] (saleContinues) at (4.6,2.8) {saleContinues};
\node[event] (setAllowance)  at (7.6,2.8) {setAllowance};
\node[event] (transferFromSuccess) at (7.6,1.0) {transferFromSuccess};
\begin{pgfonlayer}{background}
\node[lane, fit=(enableSale)(disableSale)(withdrawEth),
      label={[lanelabel]north west:owner}] {};
\node[lane, fit=(transferFail)(saleContinues)(setAllowance)(transferFromSuccess),
      label={[lanelabel]north west:any}] {};
\end{pgfonlayer}
\draw[relincl] (enableSale) -- (saleContinues)
    node[pos=0.5, above right=-2pt, font=\normalsize, text=dcrinclusion] {$+$};
\draw[relexcl] (disableSale) -- (saleContinues)
    node[pos=0.5, right=1pt, font=\normalsize, text=dcrexclusion] {$\%$};
\draw[relresp] (saleContinues) -- (withdrawEth)
    node[pos=0.58, guard, below=2pt, sloped, text=dcrresponse] {balance $>$ 0};
\draw[relcond] (setAllowance) -- (transferFromSuccess)
    node[midway, guard, left=2pt, text=dcrcondition] {allowance $\geq$ amt};
\end{tikzpicture}%
}

\vspace{5pt}
\begin{tabular}{@{}l@{\hspace{2em}}l@{}}
\tikz[baseline=-0.6ex]\draw[relcond] (0,0) -- (0.7,0);~condition &
\tikz[baseline=-0.6ex]\draw[relresp] (0,0) -- (0.7,0);~response \\[2pt]
\tikz[baseline=-0.6ex]\draw[relincl] (0,0) -- (0.7,0) node[midway, above=-1pt, font=\scriptsize, text=dcrinclusion] {$+$};~inclusion &
\tikz[baseline=-0.6ex]\draw[relexcl] (0,0) -- (0.7,0) node[midway, above=-1pt, font=\scriptsize, text=dcrexclusion] {$\%$};~exclusion
\end{tabular}

\caption{\textbf{Part of the DCR graph extracted for \textsc{Gifto}.} Dashed lanes group events by caller role; the dotted node is an isolated event; italic labels on relations are guards. The subgraph exhibits all four relation types.}
\label{fig:dcr-example}
\end{figure}

%% file: sections/evaluation.tex

\section{Evaluation}
\label{sec:evaluation}

We apply the framework to the 250-contract benchmark of \S\ref{sec:benchmark}, regenerated by open-weight LLMs spanning several families and a range of sizes, so that the findings are not tied to a single model or vendor.
After the experimental setup (\S\ref{sec:eval-setup}), we report compilation rates (\S\ref{sec:eval-compile}) and the results of differential fuzzing (\S\ref{sec:eval-fuzzing}); we then assess the quality of the extracted DCR graphs, on which O3 depends (\S\ref{sec:eval-human}), compare the two replay oracles (\S\ref{sec:eval-oracles}), and explain their relationship through the functions each of them reaches (\S\ref{sec:eval-coverage}).

\subsection{Experimental Setup}
\label{sec:eval-setup}

\paragraph{Models}
We evaluate seven models from five families: \textsc{Gemma-4-31B} and \textsc{Gemma-4-12B} (Google, instruction-tuned), \textsc{Qwen3.6-27B} and \textsc{Qwen3-Coder-30B} (Alibaba; the latter a mixture-of-experts model with 3B active parameters, specialised for code), \textsc{Ornith-35B}, \textsc{Codestral-22B} and \textsc{Devstral-24B} (Mistral, both specialised for code).
All run as 4-bit GGUF quantisations served by llama.cpp on a single RTX 5090, with a 60k-token context; generation uses temperature $0.7$ and at most $2{,}048$ output tokens per function.
Each model plays two roles: it regenerates every benchmark contract with $k = 10$ samples (\S\ref{sec:codegen}) and it extracts the DCR graph of every original contract (\S\ref{sec:dcr-formalism}).

\paragraph{Repair}
A regenerated sample that fails to compile is given one repair attempt by the same model that produced it: the model receives the sample and the compiler error, with the instruction to fix only the reported error, at temperature $0.2$.
The protocol is identical for every model; samples that still fail to compile are excluded from behavioural evaluation.

\paragraph{Oracles}
Against every compiling sample we run O1 with the configuration of \S\ref{sec:o1}, and O2 and O3 with the same budget of $1{,}000$ transactions per contract: the first ones recorded on chain for O2, and the successful calls collected by Algorithm~\ref{alg:seqgen} with $T$ equal to that budget for O3.
Since every model extracts graphs, O3 can be run either with the graphs of the model under test or with those of a single \emph{reference extractor}; which model plays this role is decided on the results, in \S\ref{sec:eval-oracles}.
All runs use seed 42.

\subsection{Compilation Separates Models but Does Not Certify Them}
\label{sec:eval-compile}

Table~\ref{tab:repair} reports Compile@10 before and after repair.
Raw rates span $18.8\%$ to $89.1\%$ and remain widely separated after repair ($26.4\%$ to $92.7\%$).
Repair lifts every model by $3.6$ to $10.8$ percentage points but \emph{preserves the ranking exactly}, and the relative gain is largest for the weakest models ($+40\%$ for \textsc{Devstral-24B}, $+4\%$ for \textsc{Gemma-4-31B}).
Whether the code that compiles is also correct is the question the remaining subsections answer.

\begin{table}[t]
  \centering\scriptsize
  \caption{\textbf{Compile@10 before and after one repair pass, per model.}\\\emph{Raw}: share of the ten samples per contract that compile as generated; \emph{Repaired}: after a single attempt in which the same model fixes its non-compiling samples from the compiler error; \emph{Gain}: their difference in percentage points.}
  \label{tab:repair}
  \setlength{\tabcolsep}{5pt}
  \begin{tabular}{@{}l r r r@{}}
    \toprule
    & \multicolumn{3}{c}{\faCode~\textbf{Compile@10}} \\
    \cmidrule(l){2-4}
    \textbf{Model} & \textbf{Raw} & \textbf{Repaired} & \textbf{Gain} \\
    \midrule
    \gemmaIcon Gemma-4-31B       & 89.1\% & \textbf{92.7\%} & \gaincell{30}{3.6}  \\
    \qwenIcon Qwen3.6-27B        & 64.7\% & \textbf{71.3\%} & \gaincell{55}{6.6}  \\
    \qwenIcon Qwen3-Coder-30B    & 53.2\% & \textbf{61.8\%} & \gaincell{72}{8.6}  \\
    \gemmaIcon Gemma-4-12B       & 48.0\% & \textbf{57.9\%} & \gaincell{82}{9.8}  \\
    \ornithIcon Ornith-35B       & 41.2\% & \textbf{52.1\%} & \gaincell{90}{10.8} \\
    \mistralIcon Codestral-22B   & 34.3\% & \textbf{39.9\%} & \gaincell{47}{5.6}  \\
    \mistralIcon Devstral-24B    & 18.8\% & \textbf{26.4\%} & \gaincell{63}{7.6}  \\
    \bottomrule
  \end{tabular}
\end{table}

\subsection{Extracted DCR Graphs Are Reliable}
\label{sec:eval-human}

O3 is only as good as the graphs it runs on, so before comparing the oracles we validated a sample of them against human judgement.
We validated the graphs of \textsc{Gemma-4-31B}, the strongest generator (Table~\ref{tab:repair}), on 30 contracts, drawn with a fixed seed by stratified sampling over three size tiers (1--5, 6--12, and 13--20 non-trivial functions) and constrained to include every relation type, since \texttt{response} is about $7\%$ of extracted relations and a plain random draw could miss it.
A single annotator, an author of this paper and a co-designer of the DCR formalism, read the Solidity source before seeing the graph, judged every extracted relation as \texttt{correct}, \texttt{incorrect}, or \texttt{unsure}, and logged the relations the extractor omitted.
Precision and recall are computed per relation type from these counts; sample, judgements, and sampling script are in the replication package.

Table~\ref{tab:human_eval} reports the results over 200 judged relations, none marked \texttt{unsure}: $0.72$ precision and $0.88$ recall overall.
\texttt{Condition}, the relation that governs ordering and therefore drives sequence generation, is the most reliable ($0.96$ precision), and \texttt{response} is close behind ($0.75$ precision, $0.95$ recall).
\texttt{Inclusion} is the weakest ($0.48$), with a systematic error: a one-time precondition, such as an allowance that must first be set, is reported as if the target event had been excluded and were now being included again, when it was only not yet enabled.
A second recurring error affects \texttt{exclusion} ($0.61$): a value driven to zero, such as an allowance fully spent, is reported as excluding a dependent event, although another function can restore the value.
Both errors add relations rather than removing them, so they over-constrain the graph; the permissive semantics of DCR and the on-chain evaluation of guards absorb part of the effect, and \S\ref{sec:eval-oracles} shows that the residual differences between extractors do not alter the results.

\begin{table}[t]
  \centering\scriptsize
  \caption{\textbf{Human validation of the DCR relations extracted by \textsc{Gemma-4-31B} on 30 contracts.} An extracted relation is a true positive (TP) if the annotator confirms it in the contract's source and a false positive (FP) otherwise; a false negative (FN) is a relation the annotator finds in the source that the extractor did not emit.}
  \label{tab:human_eval}
  \setlength{\tabcolsep}{4pt}
  \setlength{\aboverulesep}{0pt}\setlength{\belowrulesep}{0pt}
  \begin{tabular}{@{}l | rrr | rrr@{}}
    \toprule
    \hstrut\textbf{Relation} & \multicolumn{3}{c|}{\hicon{\faUserCheck}\textbf{judgements}} & \textbf{P} & \textbf{R} & \textbf{F1} \\
    \hstrut & TP & FP & FN & & & \\
    \midrule
    condition $A \dcrcondition B$  & 65 & 3  & 12 & 0.96 & 0.84  & 0.90 \\
    response  $A \dcrresponse B$   & 18 & 6  & 1  & 0.75  & 0.95  & 0.84  \\
    inclusion $A \dcrinclusion B$  & 21 & 23 & 0  & 0.48  & 1.00 & 0.65  \\
    exclusion $A \dcrexclusion B$  & 39 & 25 & 6  & 0.61  & 0.87  & 0.72  \\
    \midrule
    \rowcolor{gaingrey!25}
    \textbf{all} & \textbf{143} & \textbf{57} & \textbf{19} & \textbf{0.72} & \textbf{0.88} & \textbf{0.79} \\
    \bottomrule
  \end{tabular}
\end{table}

\subsection{The Two Replay Oracles Are Complementary}
\label{sec:eval-oracles}

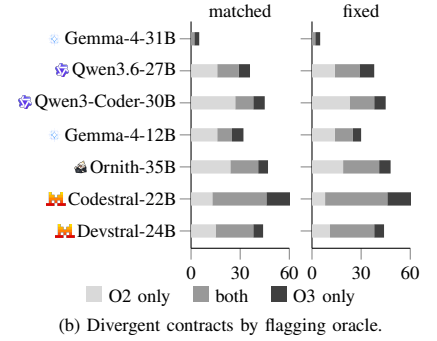
\begin{figure*}[t]
\centering
\begin{minipage}[c]{0.665\linewidth}
  \centering\scriptsize
  \setlength{\tabcolsep}{2.1pt}
  \renewcommand{\arraystretch}{1.15}
  \setlength{\aboverulesep}{0pt}\setlength{\belowrulesep}{0pt}
  \begin{tabular}{@{}l | c | rrr | rrr rrr@{}}
    \toprule
    \hstrut\textbf{Model} & \hicon{\faCode} & \multicolumn{3}{c|}{\hicon{\faHistory}\textbf{O2}} & \multicolumn{6}{c}{\hicon{\faProjectDiagram}\textbf{O3}} \\
    \cmidrule{6-11}
    \hstrut & & & & & \multicolumn{3}{c}{matched} & \multicolumn{3}{c}{fixed} \\
    \cmidrule{6-8}\cmidrule{9-11}
    \hstrut & Compile@10 & $N$ & Status & \divhdr & $N$ & Status & \divhdr & $N$ & Status & \divhdr \\
    \midrule
    \gemmaIcon Gemma-4-31B     & 92.7\% & 234 & 99.6\% & \gaincell{4}{2}    & 174 & 98.4\% & \gaincell{10}{5}   & 174 & 98.4\% & \gaincell{9}{5}    \\
    \qwenIcon Qwen3.6-27B      & 71.3\% & 214 & 96.9\% & \gaincell{63}{29}  & 157 & 96.7\% & \gaincell{40}{20}  & 162 & 95.8\% & \gaincell{45}{24}  \\
    \qwenIcon Qwen3-Coder-30B  & 61.8\% & 183 & 95.5\% & \gaincell{83}{38}  & 125 & 96.1\% & \gaincell{36}{18}  & 136 & 95.5\% & \gaincell{42}{22}  \\
    \gemmaIcon Gemma-4-12B     & 57.9\% & 169 & 95.4\% & \gaincell{54}{25}  & 127 & 94.6\% & \gaincell{32}{16}  & 129 & 94.6\% & \gaincell{30}{16}  \\
    \ornithIcon Ornith-35B     & 52.1\% & 193 & 94.4\% & \gaincell{89}{41}  & 140 & 96.1\% & \gaincell{46}{23}  & 147 & 95.2\% & \gaincell{55}{29}  \\
    \mistralIcon Codestral-22B & 39.9\% & 159 & 91.1\% & \gaincell{100}{46} & 115 & 81.8\% & \gaincell{100}{50} & 123 & 81.9\% & \gaincell{100}{53} \\
    \mistralIcon Devstral-24B  & 26.4\% & 122 & 88.2\% & \gaincell{83}{38}  &  85 & 85.5\% & \gaincell{58}{29}  &  89 & 83.3\% & \gaincell{62}{33}  \\
    \bottomrule
  \end{tabular}

  \vspace{3pt}
  (a) Per-oracle results.
\end{minipage}%
\hfill
\begin{minipage}[c]{0.325\linewidth}
  \centering\scriptsize
  \begin{tikzpicture}
  \begin{groupplot}[
    group style={group size=2 by 1, horizontal sep=0.3cm, yticklabels at=edge left},
    xbar stacked, /pgf/bar width=5pt, scale only axis, width=1.3cm, height=3.0cm,
    xmin=0, xmax=60, xtick={0,30,60},
    symbolic y coords={Devstral-24B,Codestral-22B,Ornith-35B,Gemma-4-12B,Qwen3-Coder-30B,Qwen3.6-27B,Gemma-4-31B},
    ytick=data,
    yticklabels={\gemmaIcon Gemma-4-31B,\qwenIcon Qwen3.6-27B,\qwenIcon Qwen3-Coder-30B,\gemmaIcon Gemma-4-12B,\ornithIcon Ornith-35B,\mistralIcon Codestral-22B,\mistralIcon Devstral-24B},
    y tick label style={font=\scriptsize, inner sep=1pt},
    x tick label style={font=\scriptsize},
    axis lines*=left, tick align=outside, enlarge y limits=0.09,
    every axis title/.style={at={(0.5,1)}, above, font=\scriptsize, yshift=-1pt},
  ]
  \nextgroupplot[title={matched}]
  \addplot[fill=segO2,   draw=none] coordinates {(0,Gemma-4-31B) (16,Qwen3.6-27B) (27,Qwen3-Coder-30B) (16,Gemma-4-12B) (24,Ornith-35B) (13,Codestral-22B) (15,Devstral-24B)};
  \addplot[fill=segBoth, draw=none] coordinates {(2,Gemma-4-31B) (13,Qwen3.6-27B) (11,Qwen3-Coder-30B) (9,Gemma-4-12B) (17,Ornith-35B) (33,Codestral-22B) (23,Devstral-24B)};
  \addplot[fill=segO3,   draw=none] coordinates {(3,Gemma-4-31B) (7,Qwen3.6-27B) (7,Qwen3-Coder-30B) (7,Gemma-4-12B) (6,Ornith-35B) (17,Codestral-22B) (6,Devstral-24B)};
  \nextgroupplot[title={fixed}]
  \addplot[fill=segO2,   draw=none] coordinates {(0,Gemma-4-31B) (14,Qwen3.6-27B) (23,Qwen3-Coder-30B) (14,Gemma-4-12B) (19,Ornith-35B) (8,Codestral-22B) (11,Devstral-24B)};
  \addplot[fill=segBoth, draw=none] coordinates {(2,Gemma-4-31B) (15,Qwen3.6-27B) (15,Qwen3-Coder-30B) (11,Gemma-4-12B) (22,Ornith-35B) (38,Codestral-22B) (27,Devstral-24B)};
  \addplot[fill=segO3,   draw=none] coordinates {(3,Gemma-4-31B) (9,Qwen3.6-27B) (7,Qwen3-Coder-30B) (5,Gemma-4-12B) (7,Ornith-35B) (15,Codestral-22B) (6,Devstral-24B)};
  \end{groupplot}
  \end{tikzpicture}

  \vspace{-2pt}
  \textcolor{segO2}{\rule{6pt}{6pt}}~O2 only
  \hspace{0.5em}
  \textcolor{segBoth}{\rule{6pt}{6pt}}~both
  \hspace{0.5em}
  \textcolor{segO3}{\rule{6pt}{6pt}}~O3 only

  \vspace{3pt}
  (b) Divergent contracts by flagging oracle.
\end{minipage}
\caption{\textbf{Compilation, historical replay (O2) and DCR-guided replay (O3) per model.} (a)~$N$: contracts the oracle could compare; \emph{Status}: mean share of replayed calls on which the regenerated contract $\hat{C}$ agrees with the original $C^*$ on success or revert; $\hat{C}\neq C^*$: contracts with at least one disagreement, shaded in proportion to the value. O3 is run with the DCR graphs extracted by the model itself (\emph{matched}) and with those extracted by \textsc{Gemma-4-31B} for every model (\emph{fixed}). (b)~Each bar splits a model's divergent contracts into those flagged by O2 only, by both oracles, and by O3 only, on matched and fixed graphs.}
\label{fig:oracle_comparison}
\end{figure*}

Figure~\ref{fig:oracle_comparison} compares O2 and O3 in two settings.
In the \emph{matched} setting, each model's regenerated contracts are replayed against sequences generated from DCR graphs extracted by that same model; in the \emph{fixed} setting, every model's contracts are replayed against the sequences of one reference extractor, for which we take \textsc{Gemma-4-31B}, the strongest generator (Table~\ref{tab:repair}) and, as \S\ref{sec:eval-coverage} shows, the extractor whose graphs reach the highest coverage.
Panel (a) reports, for each oracle, how many contracts it compared, how closely they agreed with the original, and how many diverged; panel (b) splits the divergent contracts by the oracle that flags them.

\paragraph{Neither oracle subsumes the other}
The decisive evidence is in panel (b).
For every model, O3 flags contracts that O2 certifies as correct (the darkest segment, from 3 contracts for \textsc{Gemma-4-31B} to 17 for \textsc{Codestral-22B}), and for every model but \textsc{Gemma-4-31B} the converse also holds (the lightest segment).
For \textsc{Qwen3.6-27B}, for instance, of the 20 contracts flagged by O3, 13 are also flagged by O2 and 7 are not, while 16 contracts diverge under O2 only.
The two oracles are therefore complementary instruments, not a stricter and a weaker version of the same test.

\paragraph{The divergences track generation quality}
Two regularities indicate that the flagged divergences are real defects rather than noise.
First, the worse a model is at producing compiling code, the more its compiling contracts diverge: under O2, the mean status match rate decreases at every step down the Compile@10 ranking, from $99.6\%$ for \textsc{Gemma-4-31B} to $88.2\%$ for \textsc{Devstral-24B}, and under O3 it decreases along the same ranking, from $98.4\%$ to $81.8\%$, with two adjacent models swapped.
Second, the two oracles order the models almost identically (Spearman $\rho = 0.86$ between their status match rates).

\paragraph{The result does not depend on the extractor}
O3 relies on an LLM-extracted graph, which raises the concern that a model could be flattered, or penalised, by the quality of its own graph.
The fixed setting removes this degree of freedom: with every model judged by the reference extractor's sequences, the number of contracts flagged by O3 only changes by at most two per model (from 3 to 17 with matched graphs, from 3 to 15 with fixed graphs; net $-1$ over the seven models) and remains non-zero for every model.
The complementarity is a property of the two transaction sources, not of which model built the graph.

\paragraph{The two oracles evaluate slightly different sets of contracts}
O2 can be applied to any contract with at least one compiling sample; O3 also needs a validated sequence for that contract (\S\ref{sec:seq-gen}), which some contracts lack.
The set of contracts O3 evaluates is therefore a subset of the set O2 evaluates, and panel (a) reports both sizes as $N$: 122 to 234 contracts for O2 and 85 to 174 for O3, depending on how many of the model's samples compile.
Hence, a contract flagged by O3 only was evaluated by both oracles, whereas a contract flagged by O2 only may be one that O3 never evaluated; \S\ref{sec:threats} quantifies this.

\paragraph{Cost}
On the runs of the reference extractor, O2 took $19.8$\,s per contract over its 234 contracts and O3 $22.1$\,s over its 174.
DCR extraction is a one-off cost per extractor, amortised over every model evaluated against its graphs.

\subsection{Coverage Explains the Complementarity}
\label{sec:eval-coverage}

The complementarity has a mechanical cause: an oracle can only expose a fault in a function it calls, and the two oracles do not call the same functions.
We say that a state-changing function of a contract is \emph{covered} by an oracle when at least one of the transactions the oracle replays calls it; state-changing functions are the only ones an oracle can exercise, and the benchmark has $1{,}759$ of them.
O2's coverage is the same for every model, since it replays the same recorded transactions; O3's depends on the graphs, hence on the extractor.
Table~\ref{tab:coverage} reports, for each extractor, the mean per-contract coverage of O3, the number of contracts on which O3 covers more functions than O2, and how many functions each oracle covers that the other does not.

\begin{table}[t]
  \centering\scriptsize
  \caption{\textbf{Function coverage of O3 per extractor, against the fixed coverage of O2.} \emph{Coverage}: mean share of a contract's state-changing functions that O3 calls at least once, with the graphs of that extractor (O2: $55.7\%$ for every row). \emph{Contracts}: contracts on which O3 covers more functions than O2. Last two columns: functions, out of $1{,}759$, that only one of the two oracles calls.}
  \label{tab:coverage}
  \setlength{\tabcolsep}{2.5pt}
  \setlength{\aboverulesep}{0pt}\setlength{\belowrulesep}{0pt}
  \begin{tabular}{@{}l | r | r | rr@{}}
    \toprule
    \hstrut\multirow{2}{*}[-2pt]{\textbf{Extractor}} & \multicolumn{1}{c|}{\multirow{2}{*}[-2pt]{\makecell{\textbf{O3}\\\textbf{coverage}}}} & \multicolumn{1}{c|}{\multirow{2}{*}[-2pt]{\makecell{\textbf{Contracts}\\\textbf{O3\,$>$\,O2}}}} & \multicolumn{2}{c}{\makecell{\textbf{Functions covered}\\\textbf{by one oracle only}}} \\
    \cmidrule{4-5}
    \hstrut & & & \hicon{\faProjectDiagram}O3 only & \hicon{\faHistory}O2 only \\
    \midrule
    \gemmaIcon Gemma-4-31B     & 47.3\% & 84 & 311 & 511 \\
    \qwenIcon Qwen3.6-27B      & 46.1\% & 81 & 305 & 520 \\
    \gemmaIcon Gemma-4-12B     & 44.6\% & 72 & 292 & 537 \\
    \mistralIcon Devstral-24B  & 44.2\% & 77 & 301 & 535 \\
    \ornithIcon Ornith-35B     & 41.6\% & 68 & 276 & 566 \\
    \mistralIcon Codestral-22B & 38.3\% & 60 & 263 & 596 \\
    \qwenIcon Qwen3-Coder-30B  & 33.6\% & 35 & 246 & 619 \\
    \bottomrule
  \end{tabular}
\end{table}

\paragraph{Each oracle reaches functions the other does not}
O2 covers $55.7\%$ of a contract's state-changing functions on average, O3 between $33.6\%$ and $47.3\%$ depending on the extractor.
The two figures are not nested.
O3 reaches functions that no historical transaction ever called, between 246 with the weakest extractor and 311 with the strongest, one state-changing function in six; these are the functions O3 was built to reach.
O2, in turn, reaches functions that O3 misses, between 619 and 511: functions whose guards the extracted arguments never satisfy, or whose relations the extractor did not emit.
A better extractor shrinks the second set but leaves the first untouched, which is why the two oracles are used together.

\paragraph{The mean hides a bimodal outcome}
The averages above conceal that the two oracles win on different contracts.
With the graphs of \textsc{Gemma-4-31B}, O3 covers more functions than O2 on 84 contracts, fewer on 121, and the same on 45; on the median contract O3 is ahead ($57.1\%$ against $50.0\%$), and only the mean favours O2.
O3 wins on contracts whose functions are gated behind preconditions that real users never satisfied, which are exactly the contracts in which ordering-dependent faults hide; O2 wins on contracts with a wide, heavily used interface, the common case in a benchmark dominated by tokens, and these dominate the mean.

\paragraph{Extracting DCR graphs and generating Solidity are different skills}
The models rank differently on the two tasks.
By Compile@10, \textsc{Devstral-24B} is the weakest generator ($26.4\%$); by the coverage its graphs give to O3, it is the third-best extractor ($44.2\%$), ahead of \textsc{Ornith-35B}, \textsc{Codestral-22B}, and \textsc{Qwen3-Coder-30B} ($33.6\%$).
Only \textsc{Gemma-4-31B} is first on both.
Since a model's graph does not reflect its generation ability, \S\ref{sec:eval-oracles} also judges every model with the same extractor's graphs.

\subsection{Differential Fuzzing Finds Few Equivalent Functions}
\label{sec:eval-fuzzing}

O1 asks a stricter question than the replay oracles: not whether a regenerated contract accepts and rejects the same transactions as the original, but whether each of its functions returns the same result on any input Echidna can find.
Of the 250 contracts, 244 expose at least one fuzzable function; the remaining six take only arguments of non-primitive type.
O1 is run on every compiling sample of these 244 contracts, one harness per fuzzable function, and Table~\ref{tab:fuzzing} reports its two measures per model.

\begin{table}[t]
  \centering\scriptsize
  \caption{\textbf{Differential fuzzing (O1) per model,} over the $N{=}244$ contracts with a compiling sample and a fuzzable function. \emph{Functions}: share of (function, sample) pairs where the regenerated function is equivalent. \emph{Contracts}: share with a sample where every fuzzable function is equivalent.}
  \label{tab:fuzzing}
  \setlength{\tabcolsep}{5pt}
  \setlength{\aboverulesep}{0pt}\setlength{\belowrulesep}{0pt}
  \begin{tabular}{@{}l | rr@{}}
    \toprule
    \hstrut\textbf{Model} & \multicolumn{2}{c}{\hicon{\faBug}\textbf{O1: equivalent}} \\
    \hstrut & Functions & Contracts \\
    \midrule
    \gemmaIcon Gemma-4-31B     & 33.4\% & 34.0\% \\
    \qwenIcon Qwen3.6-27B      & 35.5\% & 34.4\% \\
    \qwenIcon Qwen3-Coder-30B  & 26.6\% & 29.9\% \\
    \gemmaIcon Gemma-4-12B     & 18.8\% & 13.1\% \\
    \ornithIcon Ornith-35B     & 16.9\% & 15.6\% \\
    \mistralIcon Codestral-22B &  9.3\% &  5.3\% \\
    \mistralIcon Devstral-24B  &  2.0\% &  2.1\% \\
    \bottomrule
  \end{tabular}
\end{table}


\paragraph{Most regenerated functions are not equivalent}
Averaged over all fuzzable (contract, function) pairs and all models, a regenerated function body is equivalent to the original in only $31.4\%$ of the compiling samples, and only $34\%$ of the contracts have at least one sample in which every fuzzable function is equivalent.
The \emph{Functions} column of Table~\ref{tab:fuzzing} ranges from $2.0\%$ for \textsc{Devstral-24B} to $35.5\%$ for \textsc{Qwen3.6-27B}.
The ranking closely tracks Compile@10 (Table~\ref{tab:repair}): five of the seven models keep their exact rank, and only the top two swap, with \textsc{Qwen3.6-27B} ($35.5\%$) narrowly ahead of \textsc{Gemma-4-31B} ($33.4\%$) despite the latter's higher compilation rate ($92.7\%$ against $71.3\%$).
Even \textsc{Gemma-4-31B}, whose samples compile most reliably, leaves two-thirds of its fuzzable functions non-equivalent to the original.
The gap between compiling and equivalent is therefore not a property of the weakest models but of all of them.


\paragraph{Agreement with the replay oracles}
O1 flags a contract as soon as one of its functions is non-equivalent in one compiling sample, the same any-sample rule under which O2 and O3 flag a contract on a status disagreement (\S\ref{sec:comparison-axes}), so the three divergent sets can be compared directly.
Table~\ref{tab:overlap} does so over the contracts that all three oracles evaluated, with O3 run on the graphs of the reference extractor introduced in \S\ref{sec:eval-oracles}.
O1 alone accounts for most of each model's divergent contracts ($38\%$--$69\%$ of $N$, the \emph{O1} column of Table~\ref{tab:overlap}), far more than O2 or O3 alone (at most one contract, for any model).
This is not a loss of signal from the replay oracles: O1's per-function budget (up to $10{,}000$ fuzzed calls) is far larger than O2/O3's fixed $1{,}000$-transaction budget, so within $N$ it catches nearly every contract they flag — across all seven models, only one contract is flagged by a replay oracle without also being flagged by O1, consistent with a multi-transaction sequence dependency that single-function fuzzing cannot compose.
The complementarity of \S\ref{sec:eval-oracles} therefore holds between O2 and O3, not between O1 and the replay oracles: within $N$, O1 is strictly the stronger test.
All three oracles agree on $1.2\%$--$30.8\%$ of contracts per model (the \emph{all} column), rising for the weaker models.
O1's advantage does not make O2 and O3 redundant: its inputs may be implausible, while O2/O3 divergences reflect realistic usage.
The oracles differ structurally too: O1 cannot compose sequences, so it misses the state-dependent fault the one replay-only contract in Table~\ref{tab:overlap} represents.
\begin{table}[t]
  \centering\scriptsize
  \caption{\textbf{Divergent contracts by the combination of oracles that flag them, per model.} Counted over the $N$ contracts evaluated by all three oracles, with O3 on the graphs of the reference extractor. A contract is flagged by O1 if at least one fuzzable function is non-equivalent in at least one compiling sample, and by O2 or O3 if at least one sample disagrees with the original on the status of at least one transaction. The seven columns partition the contracts flagged by \emph{at least one} oracle and therefore sum to at most $N$; the remainder agreed with the original under every oracle.}
  \label{tab:overlap}
  \setlength{\tabcolsep}{3pt}
  \setlength{\aboverulesep}{0pt}\setlength{\belowrulesep}{0pt}
  \begin{tabular}{@{}l | r | rrr | rrr | r@{}}
    \toprule
    \hstrut\textbf{Model} & $N$ & \multicolumn{3}{c|}{\textbf{one oracle only}} & \multicolumn{3}{c|}{\textbf{two oracles}} & \textbf{all} \\
    \hstrut & & \hicon{\faBug}O1 & \hicon{\faHistory}O2 & \hicon{\faProjectDiagram}O3 & O1,O2 & O1,O3 & O2,O3 & \\
    \midrule
    \gemmaIcon Gemma-4-31B     & 170 & 93 & 0 & 0 &  0 &  3 & 0 &  2 \\
    \qwenIcon Qwen3.6-27B      & 158 & 70 & 0 & 0 &  5 &  8 & 0 & 15 \\
    \qwenIcon Qwen3-Coder-30B  & 133 & 50 & 0 & 0 & 15 &  7 & 0 & 14 \\
    \gemmaIcon Gemma-4-12B     & 126 & 87 & 0 & 0 & 10 &  5 & 0 & 11 \\
    \ornithIcon Ornith-35B     & 143 & 99 & 0 & 0 &  7 &  6 & 0 & 21 \\
    \mistralIcon Codestral-22B & 120 & 59 & 0 & 0 &  7 & 15 & 0 & 37 \\
    \mistralIcon Devstral-24B  &  85 & 46 & 1 & 0 &  4 &  6 & 0 & 26 \\
    \bottomrule
  \end{tabular}
\end{table}

%% file: sections/threats_to_validity.tex
\section{Threats to Validity}
\label{sec:threats}

\paragraph{Internal validity}
O3 depends on an LLM-extracted graph: an incomplete graph lowers coverage, an over-constrained one blocks valid sequences.
Three safeguards bound the effect.
DCR's permissive semantics make a missing relation cost coverage rather than correctness, since an unconstrained event is simply callable at any time; every guard is evaluated on the deployed contract, so a wrong guard can only prune a call, never admit one the original rejects; and the fixed-extractor experiment of \S\ref{sec:eval-oracles} replaces every model's graphs by another model's and moves the number of contracts flagged by O3 only by at most two per model, a perturbation far larger than any residual extraction error.

The human validation of \S\ref{sec:eval-human} rests on a single annotator.
Judging 200 relations and logging omissions requires reading the full source of 30 contracts with up to 20 functions each and tracing every \texttt{require} and branch, about 30 minutes per contract, or roughly 15 hours of expert time; a second annotator qualified to judge DCR semantics would have to be recruited from the handful of people who master the formalism, and the one we have is one of its designers.
The judgement is also not a matter of taste: each relation is checked against a precondition that either is or is not in the source, and the two error patterns found (\texttt{inclusion} mistaken for \texttt{condition}, reversible zeroing mistaken for \texttt{exclusion}) are systematic, not idiosyncratic.
The sample and the judgements are released, so the validation can be re-annotated independently.

Two design choices make our divergence counts lower bounds.
We decide divergence on transaction status, not on the revert reason or on return values, which would require tracing each of the roughly 17 million replayed calls ($250$ contracts $\times$ $10$ samples $\times$ $1{,}000$ transactions $\times$ $7$ models, per oracle); two samples that revert for different reasons therefore count as agreeing, so every divergence we report is real and some are missed.
And O3 replays only calls that the original accepts, so it cannot catch a regenerated contract that accepts a call the original rejects; replaying the pruned calls as a negative test would attribute to the regenerated contract every failure of our own argument values, which is why we exclude it, but the $44{,}019$ pruned calls are logged and the extension is mechanical.

Repair is performed by the model under test on its own output.
A fixed external repairer would inject its own capability into every model's score; self-repair keeps each model responsible for its own code, mirrors how a developer uses one assistant, and, as Table~\ref{tab:repair} shows, leaves the ranking of the models unchanged, so no conclusion depends on it.

\paragraph{External validity}
The benchmark contains Etherscan-verified, non-proxy contracts with 1 to 20 non-trivial functions and is dominated by token-like contracts.
That is the population of verified deployments, and the framework itself is contract-agnostic; whether the complementarity of the oracles holds on multi-step DeFi, bridges, or governance code, where ordering-dependent faults matter most, is the open question we consider most worth pursuing.

The 20-function bound is not a convenience but the result of an ablation (Table~\ref{tab:funcbins}): we ran the full pipeline, repair included, on contracts binned by function count, 360 samples per bin and model.
Compile@10 after repair decreases from $72.4\%$ (1--5) to $40.8\%$ (16--20), then drops to $27.9\%$ at 21--25 and below $13\%$ beyond, where repair stops compensating.
Two effects make the cut necessary rather than merely convenient.
Generation itself does not slow down with contract size, about one second per function body in every bin, but larger contracts need more attempts to yield a compiling sample, so the cost of one usable sample rises from 4\,s (1--5) to 41\,s (16--20) and 502\,s (26--30): beyond the bound, most of the compute buys samples the oracles cannot evaluate.
And past 25 functions the models stop being distinguishable: at 26--30, six of the seven compile exactly 20 of 360 samples and the seventh 26, whereas at 16--20 the strongest and weakest are still four times apart ($70.6\%$ against $16.7\%$).
Extending the benchmark upward would therefore add contracts on which compilation is too rare to evaluate and too uniform to discriminate.

\begin{table}[t]
  \centering\scriptsize
  \caption{\textbf{Compile rate and generation cost by contract size.} \emph{Mean} and \emph{Range}: Compile@10 after repair, averaged over the seven models and between the worst and the best of them. \emph{s/sample}: seconds of generation per compiling sample. The dashed line is the benchmark's 20-function bound.}
  \label{tab:funcbins}
  \setlength{\tabcolsep}{5pt}
  \setlength{\aboverulesep}{0pt}\setlength{\belowrulesep}{0pt}
  \begin{tabular}{@{}c | r r | r@{}}
    \toprule
    \hstrut\makecell{\textbf{Functions}\\\textbf{per contract}} & \multicolumn{2}{c|}{\hicon{\faCode}\textbf{Compile@10}} & \multicolumn{1}{c}{\hicon{\faClock}\textbf{s/sample}} \\
    \hstrut & Mean & Range & \\
    \midrule
    1--5   & \gaincell{100}{72.4\%} & 54.2--88.9\% & \gaincell{1}{4}     \\
    6--10  & \gaincell{68}{49.0\%}  & 35.0--69.4\% & \gaincell{3}{13}    \\
    11--15 & \gaincell{60}{43.3\%}  & 24.2--64.2\% & \gaincell{5}{24}    \\
    16--20 & \gaincell{56}{40.8\%}  & 16.7--70.6\% & \gaincell{8}{41}    \\
    \hdashline
    21--25 & \gaincell{39}{27.9\%}  &  3.9--46.4\% & \gaincell{15}{75}   \\
    26--30 & \gaincell{8}{5.8\%}    &  5.6--7.2\%  & \gaincell{100}{502} \\
    31--35 & \gaincell{11}{7.9\%}   &  2.8--14.4\% & \gaincell{83}{415}  \\
    36--40 & \gaincell{17}{12.4\%}  &  2.8--22.8\% & \gaincell{62}{310}  \\
    \bottomrule
  \end{tabular}
\end{table}

Fresh deployment starts every contract from its constructor state rather than from accumulated mainnet state, so a fault that manifests only late in a contract's life can escape; in exchange, initialisation and access-control faults are detected reliably and the benchmark runs without an archival node.
All models run as 4-bit quantisations with one temperature and $k = 10$ samples, identical for every model, so the comparison is fair even if absolute rates would shift under other settings; the selection probabilities $\alpha$ and $\beta$ were not varied, but the fixed-extractor control, which perturbs the sequences far more than any plausible change of $\alpha$ or $\beta$, left the results unchanged.

\paragraph{Construct validity}
We measure behavioural equivalence on the exercised transactions, not functional correctness.
A regenerated contract that agrees with the original on every replayed call may still differ on inputs no oracle produced, which is precisely why O1 searches for distinguishing inputs instead of replaying a fixed set, and why the three oracles are used together.
The original contract is our only specification, so a regenerated contract that fixes a fault of the original is scored as divergent; the admission filter of \S\ref{sec:benchmark}, requiring each original to reproduce at least $80\%$ of its own history, guarantees that the reference is reproducible, not that it is correct, and any such case counts against the regenerated contract, in the conservative direction.

%% file: sections/conclusions.tex
\section{Conclusion}
\label{sec:conclusion}

A regenerated smart contract has no specification of correct behaviour other than the original it was regenerated from, so the original is the only available oracle.
On this basis we built \acronym, a benchmark of 250 reproducibility-verified mainnet contracts, and evaluated seven open-weight models with three behavioural oracles: differential fuzzing, historical transaction replay, and, our core contribution, replay of synthetic transactions generated from a DCR graph of the contract's intended lifecycle.

Three findings stand out.
Compilation is a poor proxy for correctness: compile rates range from $26.4\%$ to $92.7\%$, yet only $31.4\%$ of the regenerated functions that compile are equivalent to the original, and for six of the seven models at least one compiling contract in eight diverges from the original under each replay oracle.
The replay oracles are complementary rather than redundant: for every model, DCR-guided replay exposes divergent contracts that historical replay certifies as correct, and for all but one model the converse also holds, whichever model extracts the graphs; the reason is coverage, since DCR-guided sequences reach one state-changing function in six that no historical transaction ever called, typically owner-only setup and administrative operations.
And LLM-extracted DCR graphs are reliable enough to drive the oracle, with $0.96$ precision on the \texttt{condition} relations that govern ordering, while extracting a faithful graph and generating correct Solidity prove to be distinct capabilities.

Three directions follow directly.
Extending the human validation to more contracts, extractors, and annotators would let extraction fidelity be correlated with oracle divergence per model.
Extending the benchmark to DeFi protocols, bridges, and governance systems would test the complementarity where ordering-dependent faults matter most.
And pairing DCR-guided replay with a fuzzer for the arguments, and with a negative-testing oracle that replays the pruned calls and checks that the regenerated contract rejects them too, would remove the two limits of the current corpus with no new machinery.

%% file: sections/bio.tex
\vspace{-33pt}

\begin{IEEEbiography}[{\raisebox{8mm}
{\includegraphics[width=2.4cm,height=2.4cm,clip,keepaspectratio]{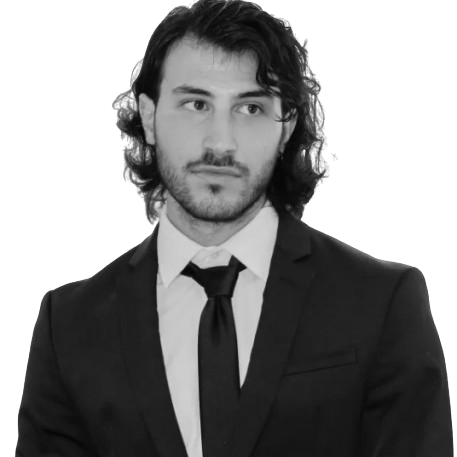}}}]
{Giovanni Pio Delvecchio} 
received the bachelor's degree in computer science from the University of Bari and the master's degree in artificial intelligence from the University of Bologna, Italy.
He is currently pursuing the PhD degree in computer science and engineering at the University of Bologna. His research interests include neuro-symbolic artificial intelligence, with a focus on its application to finance, law, and security.
\end{IEEEbiography}

\vspace{-33pt}

\begin{IEEEbiography}[{\raisebox{8mm}{\includegraphics[width=2.4cm,height=2.4cm,clip,keepaspectratio]{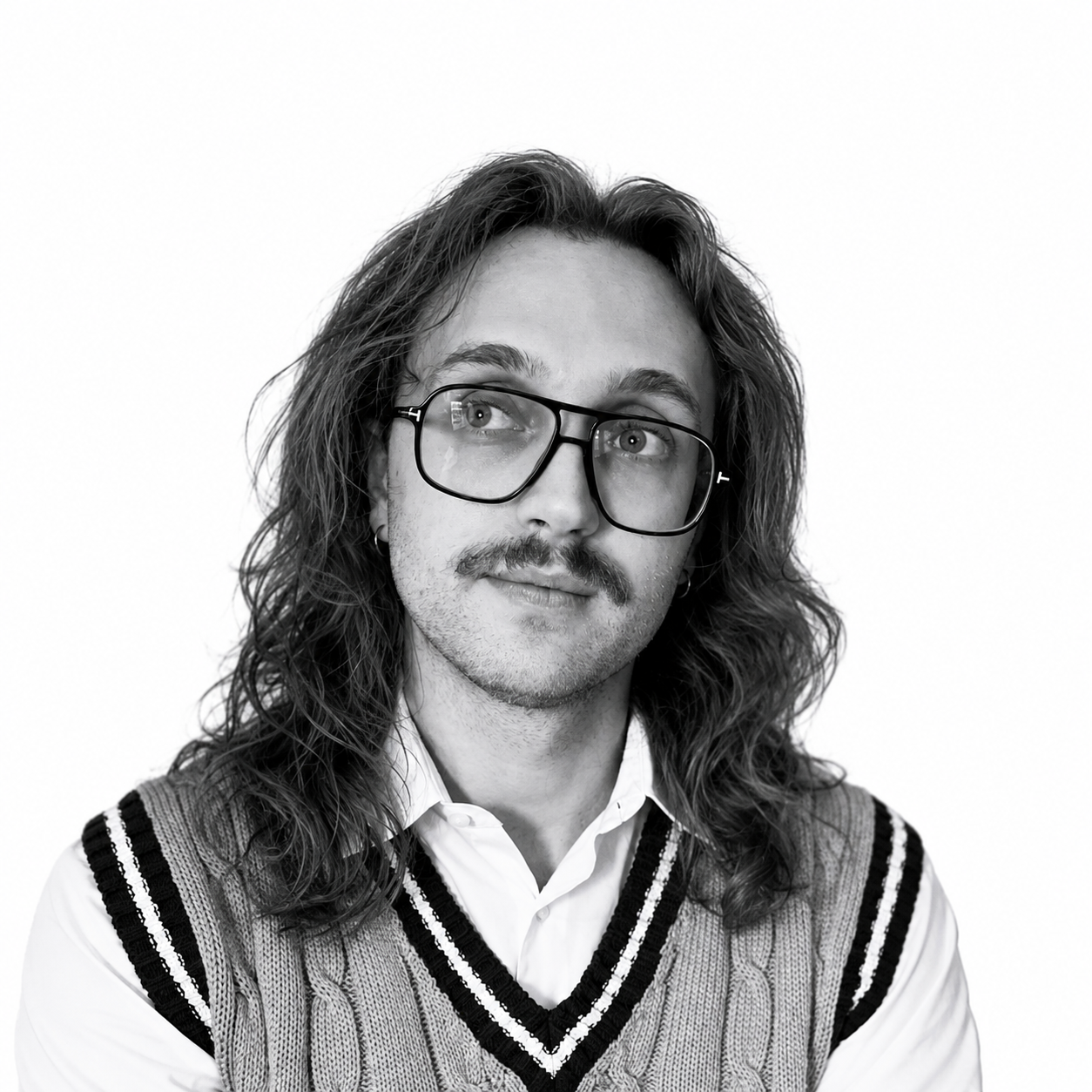}}}]{Stefano Fantazzini} received the bachelor's degree in computer engineering and the master's degree in artificial intelligence from the University of Bologna, Italy. He is currently pursuing the PhD degree in computer science and engineering at the University of Bologna. His research interests include automatic prompt learning, with a particular focus on soft prompt optimization, aiming to improve adaptation, efficiency, and reasoning across domains.
\end{IEEEbiography}

\vspace{-33pt}

\begin{IEEEbiography}[{\raisebox{8mm}{\includegraphics[width=2.4cm,height=2.4cm,clip,keepaspectratio]{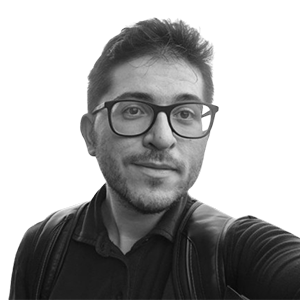}}}]{Giacomo Frisoni} received the PhD degree in computer science and engineering from the University of Bologna, Italy. He is a postdoctoral researcher in the same institution. 
His research interests include language models and graph neural networks. He has co-authored more than 30 publications and received best-paper and reviewer awards.
\end{IEEEbiography}

\vspace{-33pt}

\begin{IEEEbiography}[{\raisebox{8mm}{\includegraphics[width=2.4cm,height=2.4cm,clip,keepaspectratio]{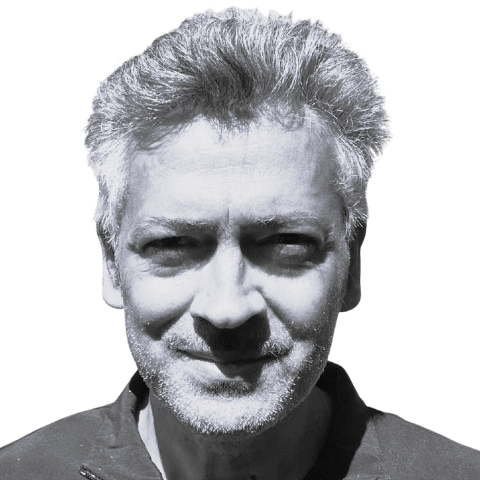}}}]{Gianluca Moro} received the PhD degree in computer engineering from the University of Bologna, Italy. 
He is an associate professor at the same institution, where he coordinates the NLP research group at the Cesena campus.
He has co-authored 100+ publications. He serves on the editorial board of multiple international journals, and collaborates on scientific projects with research centers and companies.
\end{IEEEbiography}

\vspace{-33pt}

\begin{IEEEbiography}[{\raisebox{8mm}{\includegraphics[width=2.4cm,height=2.4cm,clip,keepaspectratio]{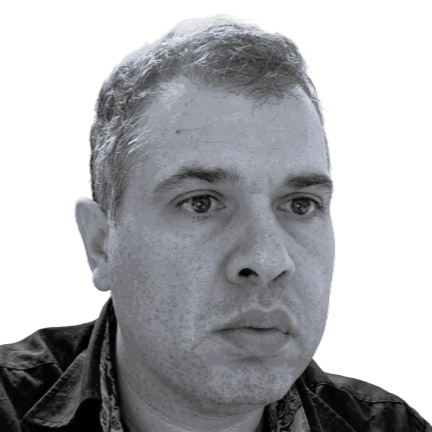}}}]{Mario Bravetti} is a Full Professor at the Department of Computer Science and Engineering of the University of Bologna. 
His research spans formal methods for distributed systems, concurrency, probability, and service-oriented and cloud computing.
He has authored more than 100 publications, serves on editorial boards, and has held program roles at CONCUR, SEFM, and SAC.
\end{IEEEbiography}

\vspace{-33pt}

\begin{IEEEbiography}[{\raisebox{8mm}{\includegraphics[width=2.4cm,height=2.4cm,clip,keepaspectratio]{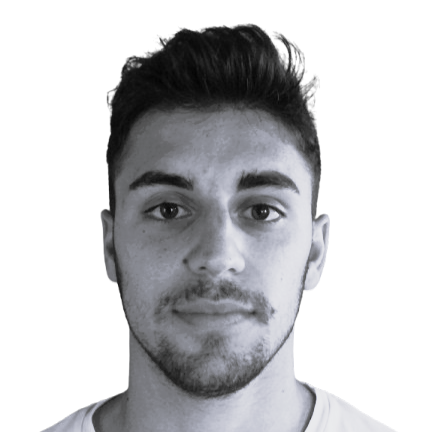}}}]{Lorenzo Bacchiani} received his PhD in computer science and engineering from the University of Bologna, Italy. He is currently a Research Fellow at the Department of Computer Science and Engineering, University of Bologna. 
His interests include behavioral methodologies for component interaction and adaptation, and digital twin technologies in the automotive domain.
\end{IEEEbiography}

\vspace{-33pt}

\begin{IEEEbiography}[{\raisebox{8mm}{\includegraphics[width=2.4cm,height=2.4cm,clip,keepaspectratio]{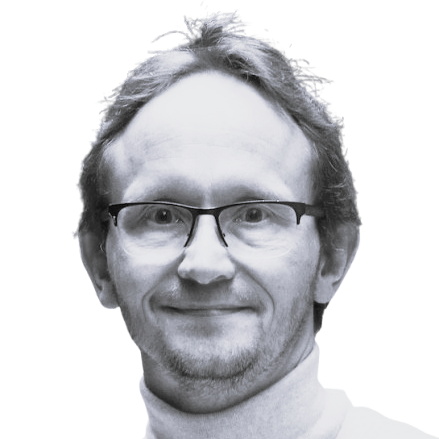}}}]{Thomas Troels Hildebrandt} received the PhD degree in semantics for concurrent systems from Aarhus University, Denmark. He is a professor in software engineering at the Department of Computer Science of the University of Copenhagen, where he heads the research section for Software, Data, People \& Society. 
For more than ten years he has led interdisciplinary projects on reliable, people-centred software systems, including the DCR graph notation.
\end{IEEEbiography}